\documentclass[iop, apj]{emulateapj}

\usepackage{epstopdf}
\usepackage{amsmath}
\usepackage{epsfig}
\usepackage[section] {placeins}

\def\deg      {{\ifmmode^\circ\else$^\circ$\fi}} 

\defcitealias{Wyder2007}{Paper I} 
\defcitealias{Martin2007}{Paper III} 
\defcitealias{Salim2007}{S07}  
 
\begin{document}

\bibliographystyle{apj}

\defcitealias{Thilker07}{T07}


 
 \title{The GALEX Arecibo SDSS Survey VII: The Bivariate Neutral Hydrogen---Stellar Mass Function for Massive Galaxies}
 
 \author{
 Jenna J. Lemonias\altaffilmark{1},
David Schiminovich\altaffilmark{1},
Barbara Catinella\altaffilmark{2},
Timothy M. Heckman\altaffilmark{3},
Sean M. Moran\altaffilmark{4}
}
 
 \altaffiltext{1}{Department of Astronomy, Columbia University, 550 West 120th Street, New York, NY 10027, USA; email: jenna@astro.columbia.edu}
 \altaffiltext{2}{Max-Planck Institut f\"{u}r Astrophysik, Karl-Schwarzschild-Str. 1, D-85741 Garching, Germany}
 \altaffiltext{3}{Department of Physics and Astronomy, The Johns Hopkins University, 3400 N. Charles Street, Baltimore, MD 21218, USA}
 \altaffiltext{4}{Smithsonian Astrophysical Observatory, 60 Garden Street, Cambridge, MA 02138, USA}
 
 
  
  

\begin{abstract}
We present the bivariate neutral atomic hydrogen (HI)---stellar mass function (HISMF) $\phi(M_{HI}, M_{*})$ for massive (log M$_{*}$/M$_{\odot}$ $>$ 10) galaxies derived from a sample of 480 local (0.025 $<$ z $<$ 0.050) galaxies observed in HI at Arecibo as part of the GALEX Arecibo SDSS Survey (GASS).   
We fit six different models to the HISMF and find that a Schechter function that extends down to a 1\% HI gas fraction, with an additional fractional contribution below that limit, is the best parametrization of the HISMF.  We calculate $\Omega_{HI, M_*>10^{10}}$ and find that massive galaxies contribute 41\% of the HI density in the local universe.  In addition to the binned HISMF we derive a continuous bivariate fit, which reveals that the Schechter parameters only vary weakly with stellar mass: M$_{HI}^*$, the characteristic HI mass, scales as M$_*^{0.39}$, $\alpha$, the slope of the HISMF at moderate HI masses, scales as M$_*^{0.07}$, and $f$, the fraction of galaxies with HI gas fraction greater than 1\%, scales as M$_*^{-0.24}$.  The variation of $f$ with stellar mass should be a strong constraint for numerical simulations.  To understand the physical mechanisms that produce the shape of the HISMF we redefine the parameters of the Schechter function as explicit functions of stellar mass and star formation rate to produce a trivariate fit.  This analysis reveals strong trends with SFR.  While M$_{HI}^*$ varies weakly with stellar mass and SFR (M$_{HI}^*$ $\propto$ M$_*^{0.22}$, M$_{HI}^*$ $\propto$ SFR$^{-0.03}$), $\alpha$ is a stronger function of both stellar mass and especially star formation rate ($\alpha$ $\propto$ M$_*^{0.47}$, $\alpha$ $\propto$ SFR$^{0.95}$).  The HISMF is a crucial tool that can be used to constrain cosmological galaxy simulations, test observational predictions of the HI content of populations of galaxies, and identify galaxies whose properties deviate from average trends.  
\end{abstract}

 
 


\section{INTRODUCTION}
\label{sec:intro}

The cold gas content of a galaxy provides insight into a galaxy's accretion history and its potential for future star formation activity.  It reflects a complex interaction between gas accretion, gas consumption, and feedback processes that inhibit both.  Recent simulations have sought to determine which processes dominate in different types of galaxies.  They have experimented with various levels and types of inflow, outflow, and feedback \citep{Hopkins2008, Keres2009, Dave2011b, Dave2011a, Leitner2011, Duffy2012, Kauffmann2012, Kim2013} and tested a range of star formation laws \citep{Lagos2011b, Lagos2011a, WangL2012} to understand the roles that each may play in determining the gas and stellar masses of galaxies.  Observations, meanwhile, have focused on constraining the global distribution of cold gas in the local universe and studying how the cold gas content of galaxies is related to their evolution and growth.

Attempts to observationally describe the distribution of galaxies in terms of their atomic gas content include HI surveys of large samples of galaxies or large areas of the sky \citep{Haynes1979, Fisher1981, Giovanelli1985} and have often relied on the measurement of HI mass functions \citep{Shostak1977, Briggs1993, Rao1993, Zwaan1997, Henning2000, Rosenberg2002, Zwaan2003, Zwaan2005, Martin2010}.  HI mass functions (HIMFs) estimate the space density of galaxies as a function of HI mass and are generally parametrized by a Schechter function \citep{Schechter1976}.  The HIMF can shed light on many aspects of galaxy evolution.  Comparisons of HIMFs derived from galaxies in different environments and across morphological types provide insight into the processes of gas stripping and accretion \citep{Solanes1996, Springob2005}.  The HIMF also provides a constraint for models of galaxy formation and evolution \citep[e.g.][]{Lagos2011b, Duffy2012, Kauffmann2012, Lu2012, Dave2013, Kim2013}.  Another motivation for constructing HIMFs was to examine the low-mass end of the HIMF for evidence of a population of ``dark" galaxies with no optical counterparts, so-called ``missing satellites" \citep{Zwaan1997, Henning2000, Rosenberg2002}.  

Early HIMFs were, by necessity, derived from relatively small samples of galaxies, sometimes as small as $\sim$60, observed as part of shallow surveys.  Much larger and deeper surveys have yielded samples of thousands of galaxies that better constrain the HIMF and can map the large-scale structure of the local universe \citep{Zwaan2003, Zwaan2005, Martin2010}.  Remarkably, the shape of the HIMF has not changed significantly since the first HIMFs based on fewer galaxies.  The large samples have, however, facilitated more complex analyses of the HI content of galaxies including the relation between HI and optical- and UV-derived properties such as stellar mass, color, and SFR \citep[e.g.][]{Toribio2011, Huang2012} and bivariate mass functions \citep{Zwaan2003}. 

Despite the utility of the HIMF, few studies have sought to explain the physical mechanisms that can shape the full distribution.  The HIMF is also hard to interpret in an evolutionary context---the origin of HI in galaxies and its evolution over time are presently quite challenging to model. Like the star formation rate (SFR) of a galaxy, HI content may increase or decrease as gas is accreted, converted to stars, cooled, ionized or heated.  This contrasts with the monotonic growth history that generally characterizes the stellar mass of galaxies.  In this respect it is natural to study the HI content of galaxies by adopting the same approach as has been used for studying the distribution and evolution of SFR vs. stellar mass in galaxies, using the \emph{bivariate} distribution, as in the color-magnitude diagram and the specific SFR-stellar mass plane \citep[e.g.][]{Wyder2007, Schiminovich2007}.  Thus we construct the bivariate $M_{HI}$-$M_\star$ HIMF, which we call the HISMF and which describes the space density of galaxies in terms of both HI mass and stellar mass.  This type of HI study necessarily assumes that all HI-detected galaxies are optically detected and benefits from samples that are complete in stellar mass.  This strategy is behind the GALEX Arecibo SDSS Survey \citep[GASS;][]{Catinella2010}, which serves as the basis of this work and is described in detail in the next section.

In this paper we use data from GASS to construct the HISMF for massive (log M$_{*}$/M$_{\odot}$ $>$ 10) galaxies, $\phi$(M$_{HI}$, M$_*$).  A key difference between this work and previous work is that we are working with a stellar mass-complete sample of galaxies, each of which has either an HI detection or upper limit.  The detections and upper limits can be used to constrain the full HISMF.  We conduct the analysis by using a Markov chain Monte Carlo (MCMC) routine to fit a Schechter \citep{Schechter1976} function and a log-normal function, and two variations of each, to the distribution of HI masses in six stellar mass bins from log M$_{*}$/M$_{\odot}$ = 10 to log M$_{*}$/M$_{\odot}$ = 11.5.  This allows us to examine for the first time how the HIMF varies with stellar mass in a large statistical and unbiased sample of local galaxies.  In addition to the binned HISMF, we also derive continuous bivariate (M$_{HI}$, M$_*$) and trivariate (M$_{HI}$, M$_*$, SFR) HIMFs.  These functions drive our discussion of the processes shaping the distribution of HI masses with respect to stellar mass.  To demonstrate the utility of the HISMF in testing simulations and observational prescriptions, we compare it to the HISMFs derived from published photometric gas fraction relations, which provide estimates of HI content based on photometrically derived quantities.

The primary goal of this work is to develop a simple parametrization of the HISMF that provides an accurate and complete description for model comparison, and that also allows some physical interpretation of its form. Recent theoretical models have already begun to predict the HIMF and interpret the physical mechanisms behind it \citep{Lagos2011b, Duffy2012, Kauffmann2012, Dave2013, Kim2013}. Our HISMF should be an important constraint for future models and will complement other recent studies seeking to build a global census of baryons in and around galaxies.  

This paper is organized as follows.  In Section 2 we describe the data and in Section 3 we describe our methodology for deriving the HISMF for massive galaxies.  We present the HISMF in Section 4.  In Section 5 we interpret the shape of the HISMF in light of the trivariate fit, and in Section 6 we summarize our findings.  The Appendix contains a comparison of our derived HISMF to the HISMF based on photometric gas fractions and an analysis of the HISMF with different assumed stellar mass functions.

\section{DATA}

The GALEX Arecibo SDSS Survey \citep[GASS;][]{Catinella2010} is an HI survey at Arecibo of $\sim$800 massive (log M$_{*}$/M$_{\odot}$ $>$ 10) local (0.025 $<$ z $<$ 0.05) galaxies.  GASS is unique for its sample selection and observing strategy.  From a parent sample of 12006 galaxies, the Arecibo targets were selected to yield a relatively flat distribution across stellar mass, ensuring uniform statistics throughout the entire stellar mass range. Our final volume statistics are determined by adopting an assumed stellar mass function.  Each galaxy was observed with Arecibo down to an HI gas \emph{fraction} limit to efficiently detect low levels of cold gas.  Because the survey is limited by gas fraction instead of HI flux, the M$_{HI}$ detection threshold varies with stellar mass.  The imposed limits are M$_{HI}$/M$_{*}$ = 1.5\% for galaxies with log M$_{*}$/M$_{\odot}$ $>$ 10.5 and a gas mass detection limit of log M$_{HI}$/M$_{\odot}$ = 8.7 for galaxies with log M$_{*}$/M$_{\odot}$ $<$ 10.5.  Since the targeted GASS sample is complete in each stellar mass bin  (the sample was selected in an unbiased way, making it representative of the true distribution of galaxies in the local universe), non-detections provide an accurate measure of the number of galaxies with gas fractions below the imposed limit.  

The sample we use is based on GASS Data Release 2 (DR2) and is identical to that described in \citet{Catinella2012}.  It contains 480 galaxies selected from the GASS parent sample, each of which is observed down to the specified limit.  GASS DR2 includes 232 Arecibo HI detections, 184 non-detections, and 64 HI detections from ALFALFA \citep{Giovanelli2005} and the Cornell HI archive \citep{Springob2005} that are added to create a statistically representative sample \citep[see][]{Catinella2010}.  Each galaxy has been observed by GALEX and SDSS, which provide homogeneously measured stellar mass and star formation rates for the sample.  Stellar masses and optical photometry are from the Max Planck Institute for Astrophysics (MPA)/Johns Hopkins University (JHU) value-added catalogs based on SDSS DR6 and DR7, respectively.  SFRs are derived from near-UV (NUV) GALEX detections and are described in detail in \citet{Schiminovich2010}.  A correction for dust attenuation is applied to star-forming galaxies with D$_{n}$(4000) $<$ 1.7 (where D$_{n}$(4000), the 4000 \AA\ break strength, is a proxy for star formation history).  HI masses for detections and HI upper limits for non-detections are calculated according to \citet{Catinella2010}.  This sample includes a small number of galaxies (N=22) whose HI masses may be contaminated by nearby gas-rich galaxies.  We found that the distribution of HI masses is nearly identical if we remove these potentially confused galaxies from the sample so we choose to keep them in our analysis \citep[see][]{Catinella2010}.

We rely on several derived quantities including the gas fraction, specific SFR, and concentration index for GASS galaxies.  We follow \citet{Catinella2010} and \citet{Schiminovich2010} in calculating these quantities.  The HI gas fraction is the HI mass divided by the stellar mass, M$_{HI}$/M$_{*}$, such that the HI gas fraction can have values above one.  We refer to this quantity alternately as the HI gas fraction or just the gas fraction, making no correction for molecular gas.  The specific SFR is the SFR normalized by stellar mass, SFR/M$_{*}$.  
The concentration index, R$_{90}$/R$_{50}$, in which R$_{90}$ and R$_{50}$ are the radii encompassing 90 and 50\% of the Petrosian flux measured in the \emph{r}-band, is used as a proxy for morphology.

\section{Methodology}
\label{sec:methodology}

\subsection{The Models}
\label{sec:models}

Astronomers have found success modeling distribution functions of galaxy properties with a \citet{Schechter1976} function.  The Schechter function has been applied to optical and near-infrared luminosity functions \citep{Bell2003_SMF}, UV luminosity functions \citep{Martin2005}, HI mass functions \citep{Zwaan2005, Martin2010}, and SFR functions \citep{Bothwell2011}.  In this paper we compare the success with which three variations of the Schechter and log-normal functions describe the HISMF.  The log-normal function is not often used to describe HIMFs, but it has been used to model other galaxy properties, including the star formation rate function \citep{Martin2005}, the X-ray luminosity function \citep{Norman2004} and the IR luminosity function \citep{Chapman2003}.  \citet{Kauffmann2003} found that the distribution of galactic size (in their case, R$_{50}$) in narrow stellar mass bins is well-fitted by a log-normal function.  Their use of the log-normal function was motivated by models showing that the spin parameter, $\lambda$, directly proportional to disk size, is distributed according to a log-normal function \citep{Fall1980, Mo1998}.  If the cold gas surface density is approximately constant across the disk of a galaxy, then it follows that the total HI mass of a galaxy should also approximate a log-normal function.  \citet{Cortese2011} also noted that HI gas fractions are distributed according to a log-normal function.

\begin{deluxetable*}{llll}  
\tablecaption{Functional Forms of the Models}
\tablehead{}
\startdata
Full Schechter & $\phi(M_{HI})\ dlogM_{HI} = \phi^{*} \ (\frac{M_{HI}}{M_{HI}^{*}})^{\alpha} \ exp({\frac{-M_{HI}}{M_{HI}^{*}}}) \ dlogM_{HI}$ & 0.01 $\times$ M$_{HI,break}$ $<$ M$_{HI}$ $<$ 10$^{12.0}$ & (1) \\
\tableline
Broken Schechter & $\phi(M_{HI}) \ dlogM_{HI} = \frac{(1-f)}{2.0} \phi^{*}$  & 0.01 $\times$ M$_{HI,break}$ $<$ M$_{HI}$ $<$ M$_{HI,break}$ & (2) \\
    & $\phi(M_{HI}) \ dlogM_{HI} = f \phi^{*} \ (\frac{M_{HI}}{M_{HI}^{*}})^{\alpha} \ exp({\frac{-M_{HI}}{M_{HI}^{*}}})$ & M$_{HI}$ $>$ M$_{HI,break}$ &  \\
\tableline
Bent Schechter & $\phi(M_{HI}) \ dM_{HI} = \phi^{*} \ (\frac{M_{HI,break}}{M_{HI}^{*}})^{\alpha} \ exp({\frac{-M_{HI,break}}{M_{HI}^{*}}})$ & 0.01 $\times$ M$_{HI,break}$ $<$ M$_{HI}$ $<$ M$_{HI,break}$ & (3) \\
    & $\phi(M_{HI}) \ dM_{HI} = \phi^{*} \ (\frac{M_{HI}}{M_{HI}^{*}})^{\alpha} \ exp({\frac{-M_{HI}}{M_{HI}^{*}}})$ & M$_{HI} > M_{HI,break}$ &  \\
\tableline
Full Log-normal & $\phi(M_{HI}) \ dlogM_{HI} = \phi^* exp({-\frac{(log M_{HI}-\mu)^2}{2 \sigma^2}}) dlogM_{HI}$ & 0.01 $\times$ M$_{HI,break}$ $<$ M$_{HI}$ $<$ 10$^{12.0}$ & (4) \\ 
\tableline
Broken Log-normal & $\phi(M_{HI}) \ dlogM_{HI} = \frac{(1-f)}{2.0} \phi^*$  & 0.01 $\times$ M$_{HI,break}$ $<$ M$_{HI}$ $<$ M$_{HI,break}$  & (5) \\
   & $\phi(M_{HI}) \ dlogM_{HI} = f \phi^* exp({-\frac{(log M_{HI}-\mu)^2}{2 \sigma^2}})$ & M$_{HI}$ $>$ M$_{HI,break}$ & \\
\tableline
Bent Log-normal & $\phi(M_{HI}) \ dM_{HI} = \phi^* exp({-\frac{(log M_{HI,break}-\mu)^2}{2 \sigma^2}})$ & 0.01 $\times$ M$_{HI,break}$ $<$ M$_{HI} < M_{HI,break}$ & (6) \\
   & $\phi(M_{HI}) \ dM_{HI} = \phi^* exp({-\frac{(log M_{HI}-\mu)^2}{2 \sigma^2}})$ & M$_{HI} > M_{HI,break}$ &  \\
\enddata
\label{tbl:models}
\end{deluxetable*}

\begin{figure}[t]
\epsscale{1.0}
\plotone{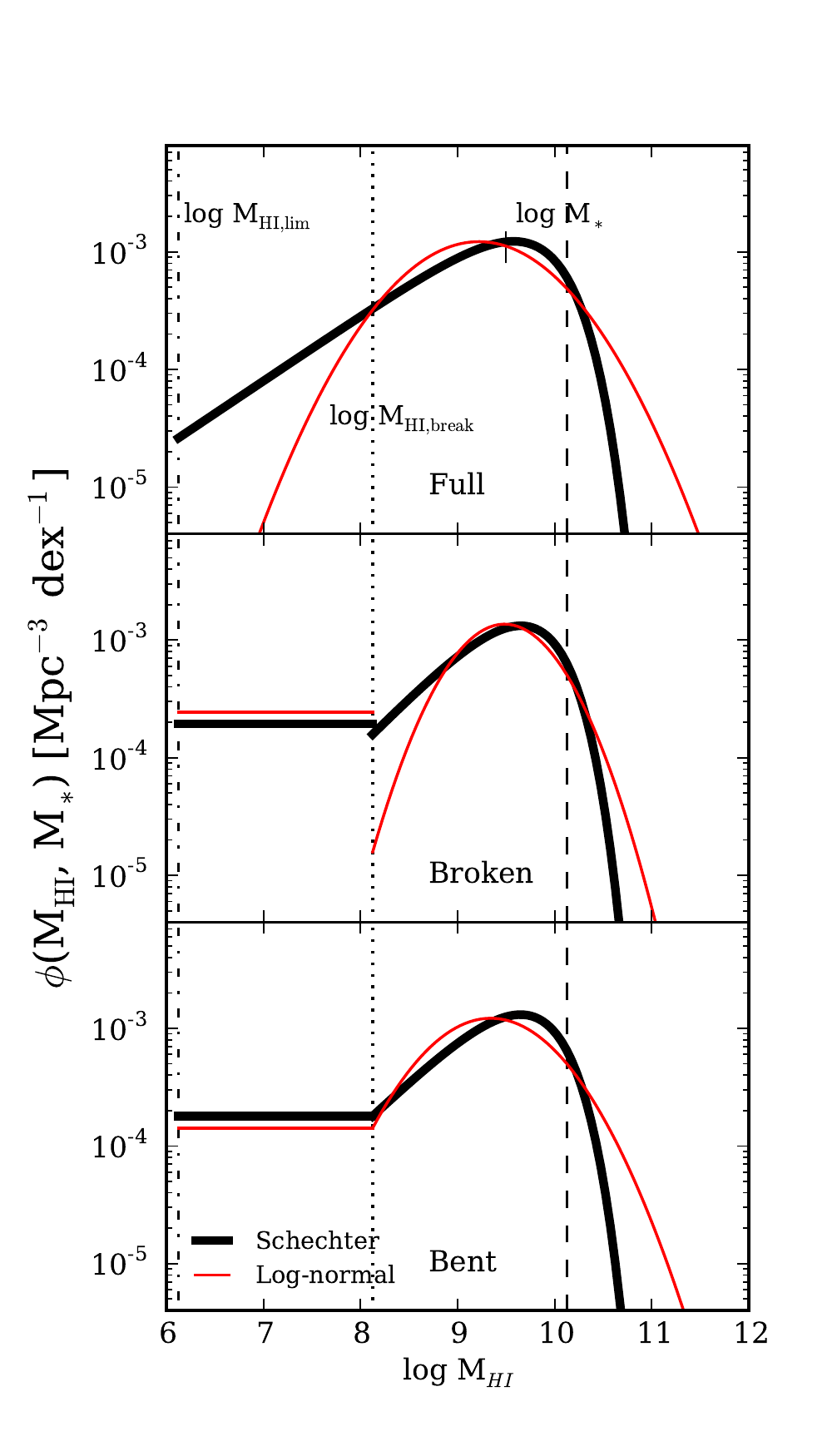}
\caption{Examples of the six models described in Section \ref{sec:models}.  From top to bottom, the panels show the full, broken, and bent variations of the Schechter function (bold black line) and log-normal function (thin red line).  Each curve shown here is a fit to galaxies with stellar masses in the range 10.0 $<$ log M$_{*}$ $<$ 10.25, as described in Section \ref{sec:himf}.} 
\label{fig:schematic} 
\end{figure} 

Here we describe the functional forms of and motivation for each model we test for the binned HISMF.  The functional forms are compiled in Table \ref{tbl:models}.  The Schechter function (Eq. 1 in Table \ref{tbl:models}) is composed of a power law that describes the distribution of low-mass galaxies and an exponential that describes the distribution of high-mass galaxies.  A schematic of the Schechter function is shown in the top panel of Fig. \ref{fig:schematic} in bold.  The power law at low HI masses ($\alpha$) characterizes an extended tail of galaxies with small amounts of HI; the exponential at high HI masses above $M_{HI}^{*}$ shows that the number of HI-rich galaxies drops off sharply above a given HI mass.   The Schechter function nicely fits the one-dimensional HIMF derived from surveys with high numbers of HI-poor galaxies \citep{Martin2010}.  However it is not apparent that the HI distribution at a given stellar mass should also be characterized by a simple power law at the HI-poor end.  In particular, for a survey such as GASS in which the tail of the distribution is dominated by non-detections below the gas fraction limit, the shape of the distribution of low-HI-mass galaxies is not well constrained.  Though upper limits provide some information about the total number of galaxies with low HI masses, we do not want upper limits to strongly constrain the shape of the function below $M_{HI}^{*}$.  

\begin{figure}[t]
\epsscale{1.0}
\plotone{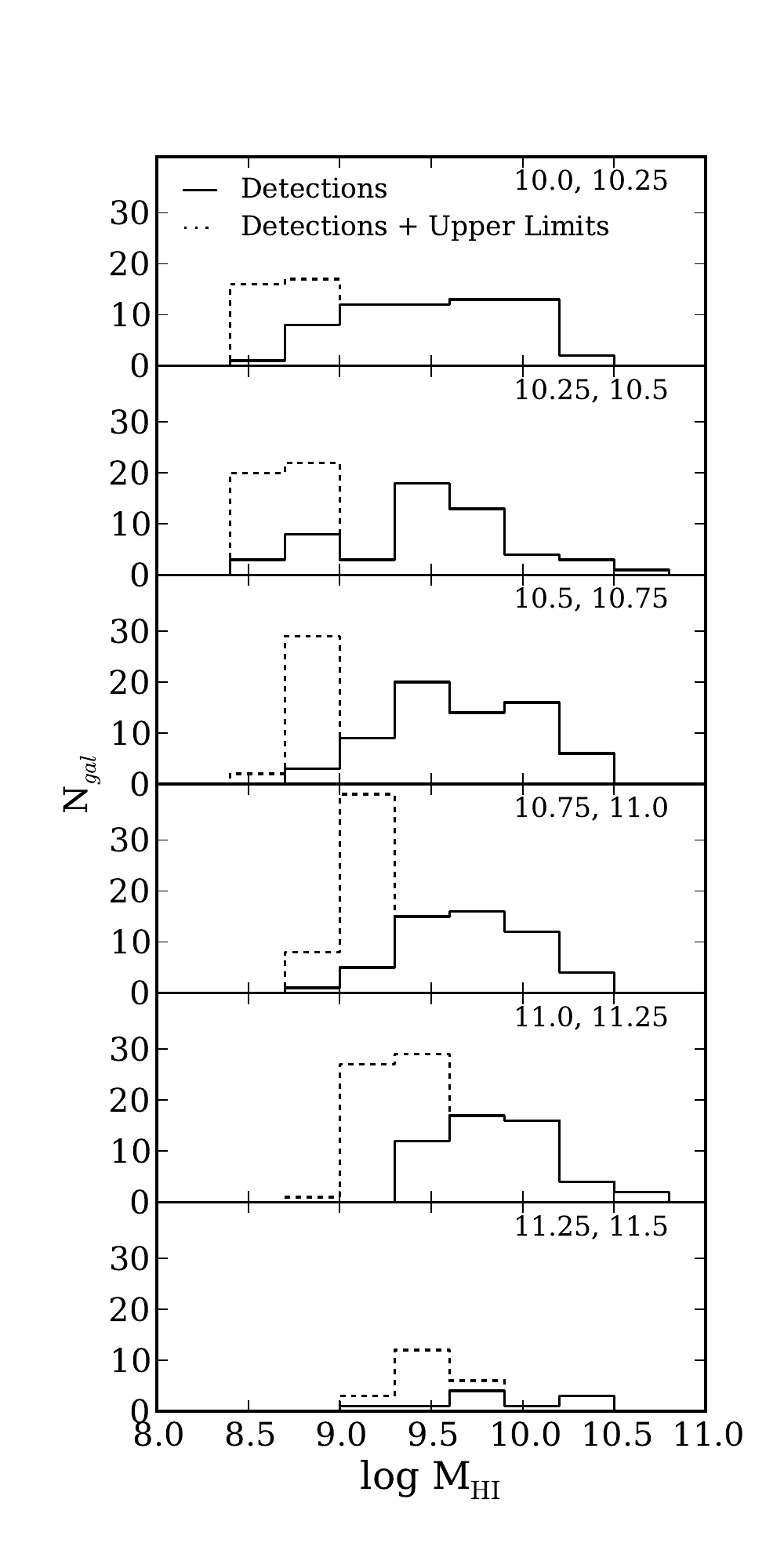}
\caption{The number of detections and upper limits per HI mass-stellar mass bin in the GASS sample.} 
\label{fig:nplot} 
\end{figure}

To account for this issue, we introduce two versions of a truncated Schechter function that do not extend to low HI masses.  In both cases the function is a combination of a Schechter function at relatively high HI masses and a constant distribution at low HI masses.  We do not necessarily expect the true distribution of HI-poor galaxies to be flat, but we choose a flat distribution for simplicity and because we have no way of determining from GASS what the true distribution is where the sample is dominated by upper limits.  The flat distribution at low HI masses reflects the fact that while we choose not to use the GASS sample to put constraints on the shape of the HISMF in this range, we can use it to derive the total number of galaxies expected to have low HI masses.  

The choice of where to impose the break in the Schechter function is somewhat arbitrary (its precise location does not significantly affect the parameters of the fit).  For simplicity we have elected to set it at a constant gas fraction with respect to stellar mass to match our observational limits.  We set M$_{HI,break}$ equal to 1\% of the central stellar mass in each mass bin:

\begin{equation}
log M_{HI,break} = log(M_{*,bin}) - 2.0,
\end{equation}

which is just below the gas fraction detection threshold for GASS.  With this choice, most HI detections are above M$_{HI,break}$ and upper limits tend to cluster around M$_{HI,break}$. 

Two variations of a Schechter function are shown in the bottom panels of Fig. \ref{fig:schematic}. 
In the middle panel the function is parametrized according to Eq. 2 in Table \ref{tbl:models}.  We call this a broken Schechter function.  In addition to $\alpha$ and M$_{HI}^{*}$, we fit another parameter, $f$, defined as the fraction of galaxies in a given stellar mass bin with HI masses above the cutoff:

\begin{equation}
f = \frac{N_{M_{HI} > M_{HI,break}}}{N_{tot}}.
\end{equation}

Because of our treatment of upper limits (see below), all upper limits contribute to the estimate of the number of galaxies below M$_{HI,break}$ even if the value of the upper limit is greater than M$_{HI,break}$.  $(1-f) \phi^{*}$ is derived so that the balance of galaxies expected to reside below M$_{HI,break}$ are uniformly contained within a range of HI masses extending up to M$_{HI,break}$.  All functions are normalized to match the \citet{Borch2006} stellar mass function. 

We must be careful to properly define the HI mass range within which we derive the HISMF.  The low HI mass cutoff impacts the probability for non-detections (see below) and the normalization of the HISMF.  The chosen lower limit should reflect the minimum HI mass expected to be associated with a galaxy of a given stellar mass, whether or not it is detectable by GASS.  We adopt a lower limit of log M$_{HI,low}$ = log M$_{*,bin}$ - 4.0  (an HI mass four dex smaller than the central stellar mass of each bin).  This limit is suggested by Ly$\alpha$ absorber measurements in \citet{Thom2012} showing that early-type and spiral galaxies in a mass range similar to that for GASS are likely to have the same minimum amount of HI in their CGM, which they give as approximately log M$_{HI}$/M$_{\odot}$ $\sim$ 6.0 and scales with galactic surface area.  This value is also generally consistent with upper limits for the least HI-rich galaxies in \citet{Serra2012}.  We set the high HI mass limit to be log M$_{HI}$ = 12.0; choosing a slightly higher or lower value would negligibly affect our results.  
 
The two main differences between the broken Schechter and the full Schechter are that the broken Schechter is discontinuous and potentially multipeaked.  We will discuss whether a multipeaked distribution function makes sense physically in Section 5.  Because the discontinuity at M$_{HI,break}$ is clearly unphysical, we introduce another function that requires the flat part of the function to connect smoothly to the Schechter part of the function.  We label this the bent Schechter function (Eq. 3 in Table \ref{tbl:models}).  An example of it is shown in the bottom panel of Fig. \ref{fig:schematic}.  Because the bent Schechter function is required to be continuous, $\alpha$ might be biased by $f$.  We will show that this constraint tends to flatten the distribution.  

Though the Schechter function has been used successfully to describe HIMFs, we also test a log-normal fit to the HISMF.  The space density of galaxies in a stellar mass bin as described by a log-normal distribution is shown as Eq. 4 in Table \ref{tbl:models}.  $\mu$ is the mean and $\sigma^{2}$ is the variance of the distribution.  A key difference between the Schechter and log-normal functions is that the log-normal function is symmetric about its mean.  We also test two variations of the log-normal function: the broken log-normal function, Eq. 5 in Table \ref{tbl:models} and Fig. \ref{fig:schematic}b, and the bent log-normal function, Eq. 6 in Table \ref{tbl:models} and Fig. \ref{fig:schematic}c.

\subsection{Continuous Fits}

\begin{deluxetable}{l}  
\tablecolumns{4}
\tablecaption{Parametrizations for Continuous Fits}
\tablehead{
\colhead{Fits Including Stellar Mass}
}
\startdata
 $\alpha = \alpha_m \times (log M_{*} -10.5) + \alpha_o$ \\
$ M_{HI}^* = M_m \times (log M_{*} - 10.5) + M_o$ \\
$ f= f_m \times (log M_{*} - 10.5) + f_o$ \\
\cutinhead{Fits Including Stellar Mass and SFR}
$ \alpha = \alpha_m \times (log M_{*} -10.5) + \alpha_s \times log SFR + \alpha_o$ \\
$ M_{HI}^* = M_m \times (log M_{*} - 10.5) + M_s \times log SFR + M_o $ \\
$ f=  f_m \times (log M_{*} - 10.5) + f_s \times log SFR + f_o $ 
\enddata
\label{tbl:otherfits_def}
\end{deluxetable}

Ultimately our goal is not just to accurately model the HISMF but to understand the physical processes that shape it.  If its shape is the result of processes linked to halo or stellar mass and SFR then one might conjecture that the parameters of the distribution function could be related to a continuous function of observable physical quantities.  Later in this paper, we examine the dependence of the HIMF on stellar mass and SFR by introducing a hierarchical model in which we redefine the distribution function parameters in terms of these physical quantities.  First we define the Schechter parameters $\alpha$, M$_{HI}^{*}$ and $f$ as functions of stellar mass and then we add SFR as an additional parameter.  The parametrizations for these hierarchical models are listed in Table \ref{tbl:otherfits}.  This fit is discussed in Section \ref{sec:otherfits}.

Determining the dependence of the HIMF on SFR can be difficult when using UV-derived SFRs because such SFRs are considered upper limits for passively evolving galaxies \citep{Schiminovich2007}.  To test that the SFRs for passively evolving galaxies do not significantly bias the fit, we implemented another version of the fit in which we treat galaxies with low specific SFRs differently.  For galaxies with log SFR/M$_*$ $<$ -12, we redefined their SFRs as a function of stellar mass only:

\begin{equation}
log SFR = log M_* - 12.0
\end{equation}

and substituted this equation for log SFR in the parametrizations in Table \ref{tbl:otherfits}.  Thus, the resulting Schechter parameters depend only on stellar mass for low star-forming galaxies.  We found that this parametrization yields results similar to the original parametrization.  When we discuss this fit in Section \ref{sec:otherfits} we only refer to the results with the original parametrization.

\subsection{Implementation}
\label{sec:implementation}

We conduct this analysis in a Bayesian framework, deriving the best-fit parameters of the models described above by maximizing the posterior probability given our dataset D = $\left\{M_{HI}, M_{HI,lim}, M_*\right\}$.  We describe the HI mass function $\phi$(M$_{HI}$) using a vector of parameters $\Theta$.  For example, $\Theta$ = $\left\{\alpha, M_{HI}^*\right\}$ for the full and bent Schechter functions and  $\Theta$ = $\left\{\alpha, M_{HI}^*, f\right\}$ for the broken Schechter function, where $\phi^\star(HI)$ is not included as a free parameter because it is derived using $\phi^\star(M_\star)$ as discussed further below.  The total posterior probability for a given set of parameters is the product of the posterior probabilities calculated for each galaxy in D: 

\begin{equation}
P(\Theta | D) = \prod\limits_{i=1}^{i=N} P_i (\Theta | D_i).
\end{equation}

We define the posterior probability for each galaxy by rewriting Bayes' theorem in accordance with our measurements and generalized model.  The posterior probability P($\Theta$ $|$ D) depends on the likelihood function of the data given the model, a prior, and a normalization factor.  It can be written as:

\begin{equation}
P_i (\Theta | D_i) = P_i (D_i | \Theta) P_i(\Theta)
\end{equation}

P$_i$($\Theta$) is the prior on the model parameters.  We impose a flat prior on $\alpha$ and $\sigma$ and a flat prior in log space on M$_{HI}^*$ and $\mu$.  We constrain $f$ to lie between 0 and 1. 

An important part of our analysis is the inclusion of upper limits for galaxies not detected in HI.  This strategy allows us to model more accurately the mass function at low HI masses and high stellar masses, where many of the non-detections lie.  We follow the method outlined in \citet{Gelman2003} for incorporation of ``censored" data.  In this method the likelihood P$_i$ (D$_i$ $|$ $\Theta$) is defined separately for galaxies detected and not detected in HI and we define an inclusion vector $I$ where $I_i$ = 1 if the galaxy is detected and $I_i$ = 0 if the galaxy is not detected in HI.  The likelihood for galaxies with HI detections is:

\begin{equation}
p (D_i | \Theta, I_i=1) \propto \phi(M_{HI,i}; \Theta),
\label{eq:det}
\end{equation}

where the proportionality is due to an extra normalization term.  The likelihood for undetected galaxies is the integral of the mass function from some lower limit to the calculated HI mass upper limit of each individual galaxy:

\begin{equation}
p (D_i | \Theta, I_i=0) \propto \int_{M_{HI,low}}^{M_{HIlim,j}} \phi(M_{HI}; \Theta) d\log M_{HI}
\label{eq:nondet}
\end{equation}

Putting everything together, we can define the total posterior of a set of parameters as:

\begin{eqnarray}
\lefteqn{ p (\Theta | D, I) = p (\Theta) p (D, I | \Theta) }\nonumber \\
   && \propto p (\Theta) \prod\limits_{i=1}^{i=N_{det}} \phi(M_{HI,i}; \Theta; I_i=1) \nonumber \\
   && \times \prod\limits_{j=1}^{j=N_{nondet}} \int_{M_{HI,low}}^{M_{HIlim,j}} \phi(M_{HI}; \Theta; I_j=0) dM_{HI}
\label{eq:total}
\end{eqnarray}

First we consider the form of the HISMF within a stellar mass bin:
$$\phi(M_{HI}) = \int_{M_\star-\Delta M_\star/2}^{M_\star+\Delta M_\star/2}\phi(M_{HI},M_\star)dM_\star$$
where $\Delta$M$_*$ is the width of the stellar mass bin and $\phi$(M$_{HI}$, M$_*$) can take any of the six forms listed above.  

We use the space density of galaxies in each stellar mass bin to determine the normalization factor for the likelihoods.  This space density is the integral of the stellar mass function $\phi(M_\star)$ within a given stellar mass bin:

\begin{equation}
\Phi(M_{\star,bin})=\int_{M_{\star,bin}-\Delta M_{\star,bin}/2}^{M_{\star,bin}+\Delta {M_\star,bin}/2}\phi(M_\star)d\log M_\star
\end{equation}

where the functional form for $\phi(M_\star)$ has been determined independently by a number of authors (see the Appendix).  With this space density we can rewrite the likelihoods with an extra normalization term $\frac{p(bin|M_{*,i})}{\Phi(M_{\star,bin})}$ where p(bin$|$M$_{*,i}$) is a step function and equals 1 when a galaxy's stellar mass is within the stellar mass bin under consideration.  Including the normalization term in Eq. \ref{eq:total} yields the exact definition of the posterior, leaving out a constant combinatorial term:

\begin{eqnarray}
\lefteqn{ p (\Theta | D, I) = p (\Theta) p (D, I | \Theta) }\nonumber \\
   && = p (\Theta) \prod\limits_{i=1}^{i=N_{det}} \phi(M_{HI,i}; \Theta; I_i=1) \frac{p(bin|M_{*,i})}{\Phi(M_{\star,bin})} \nonumber \\
   && \times \prod\limits_{j=1}^{j=N_{nondet}} \int_{M_{HI,low}}^{M_{HIlim,j}} \phi(M_{HI}; \Theta; I_j=0) \frac{p(bin|M_{*,i})}{\Phi(M_{\star,bin})} dM_{HI}
\end{eqnarray}

\subsection{Model Fitting}

We determine the parameters of the HISMF and their errors by using a Python implementation of the Markov chain Monte Carlo (MCMC) method called \emph{emcee} \citep{ForemanMackey2013}.  One of the advantages of \emph{emcee} is that it evolves an ensemble of walkers simultaneously, significantly shortening the autocorrelation time.  We run emcee using 50 walkers, 100 burn-in steps, and then 100 steps per walker.  \emph{emcee} uses the value of the posterior probability calculated for each set of parameters $\Theta$ to determine how the walkers will subsequently move in parameter space.  The final output of \emph{emcee} is a list of 50 $\times$ 100 = 5000 sets of parameters that describe the locations of the walkers in parameter space.  Because the most likely regions of parameter space are more densely occupied by the walkers, we simply take the median value of each parameter as the best-fit parameter.  Choosing the best-fit parameters by selecting the iteration with the maximum probability yields virtually identical fits, though taking the median means the best-fit parameters are more robust to artificial spikes in the probability.

We run emcee six times; the only difference between each run is the definition of the PDF for each of our six models.  In practice, and in order to treat each model identically, we express each model PDF as a finely resolved histogram.  The probability for a detection is defined as the value of the PDF at the HI mass of the detection (Eq. \ref{eq:det}).  The probability for an upper limit is defined as the integral of the PDF from some lower limit (M$_{HI,low}$) to the HI upper limit of the non-detection (Eq. \ref{eq:nondet}).  The total PDF is the product of the probabilities of each detection or upper limit.  In practice, we sum the logarithm of the probabilities.

The only source of uncertainty that we include in our model is the uncertainty on the HI measurements.  We convolve each HI measurement with its total error, which we define as the sum in quadrature of the error on the distance (which we assume to be 5\%) and the error on the HI flux.  For upper limits of GASS non-detections we set the total error to 0.3 dex.  The probability for each convolved HI measurement is the final probability that contributes to the model.  Cosmic variance could produce additional uncertainty in the normalization of the HISMF, but our sample is drawn from a region ~3 Mpc$^3$ in volume for which cosmic variance should be small \citep{Somerville2004}.

We tested our methodology by applying the same analysis to a simulated sample of galaxies with stellar masses and redshifts assigned to mimic the observed GASS sample and with HI masses extracted from a known Schechter function.  We ``observed" the simulated dataset based on the GASS detection limits, producing sets of HI masses and HI upper limits.  Checks show that the simulated sample matches the stellar mass distribution and detection fraction of the actual sample, as expected.  Applying the methodology described above on the simulated sample recovered the original Schechter function that described the simulated sample, confirming the validity of our method.  To test the effect of the imposed gas fraction limit on GASS, which becomes an HI mass limit at log M$_{*}$/M$_{\odot}$ $<$ 10.5, we also ``observed" the simulated sample with the gas fraction limit extending to different stellar masses.  This change can moderately affect $\alpha$ because it changes the proportion of detections to non-detections at low HI masses and low-to-moderate stellar masses.  We also used the simulation to test the weighting of each stellar mass bin in the continuous fits.  The relatively flat stellar mass distribution of the sample (except for a deficit of galaxies in the highest stellar mass bin) means that for the continuous fit, galaxies do not contribute to the fit in proportion to the z=0 stellar mass function.  However, we found that weighting the continuous fit by the observed stellar mass function in order to account for the stellar mass distribution of the sample does not significantly change the results; the results are robust to different weighting schemes.

\section{The Bivariate HI Mass Function for Massive Galaxies}
\label{sec:himf}

\subsection{Model Comparison}

\begin{figure*}[t]
\epsscale{1.2}
\plotone{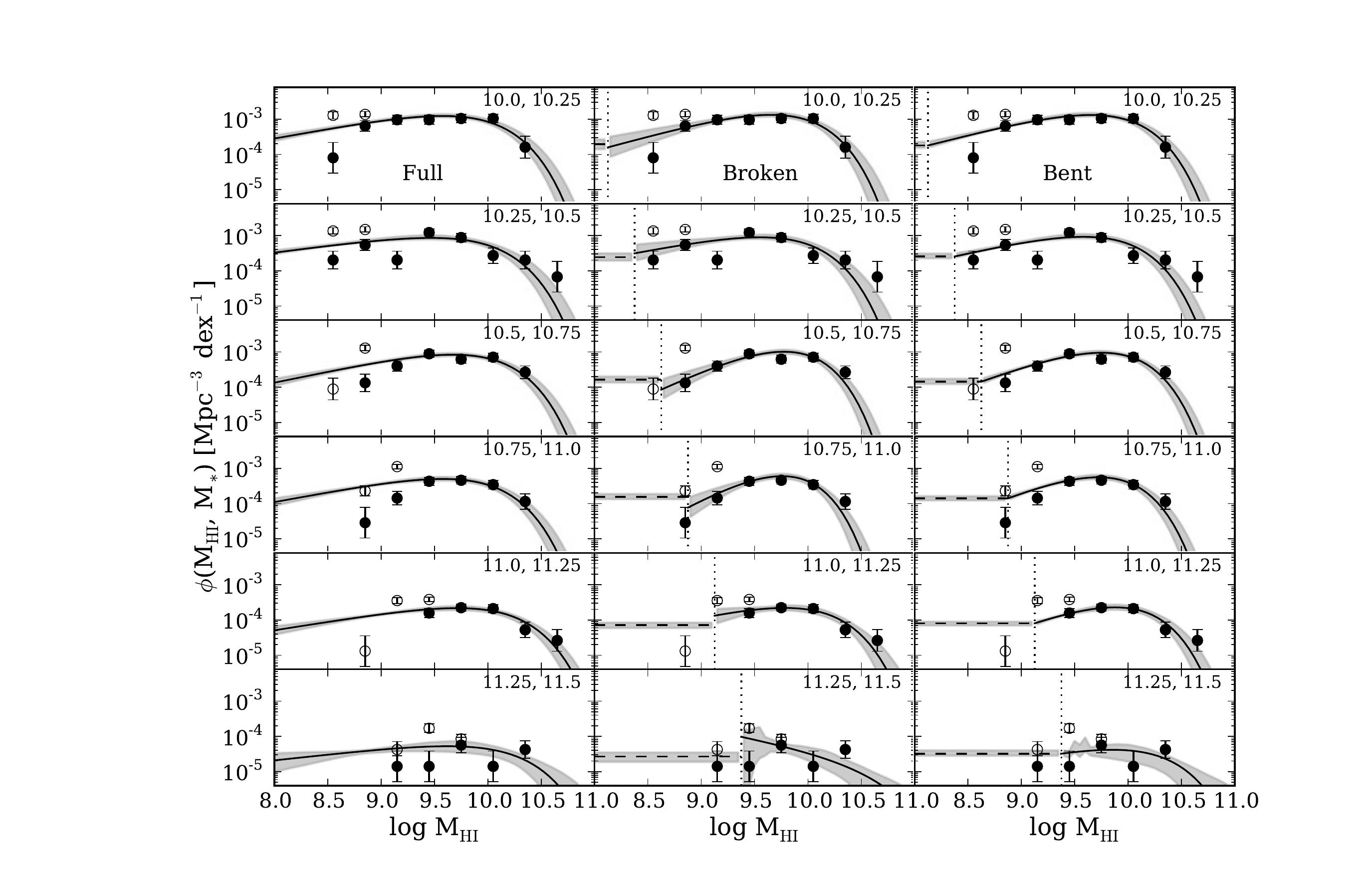}
\caption{Three variations of the Schechter function fit to the HISMF in six stellar mass bins.  The stellar mass bins are indicated in the upper right corners.  Each panel presents the binned GASS data for detections (solid black circles) and detections plus upper limits (empty circles).  Error bars on the  show Poisson uncertainties.  Shaded regions show the 1$\sigma$ uncertainties determined from the MCMC fit.  Vertical dotted lines indicate M$_{HI,break}$ = 0.01 $\times$ M$_{*,bin}$ in each stellar mass bin.}
\label{fig:schechter} 
\end{figure*}

\begin{figure*}[t]
\epsscale{1.2}
\plotone{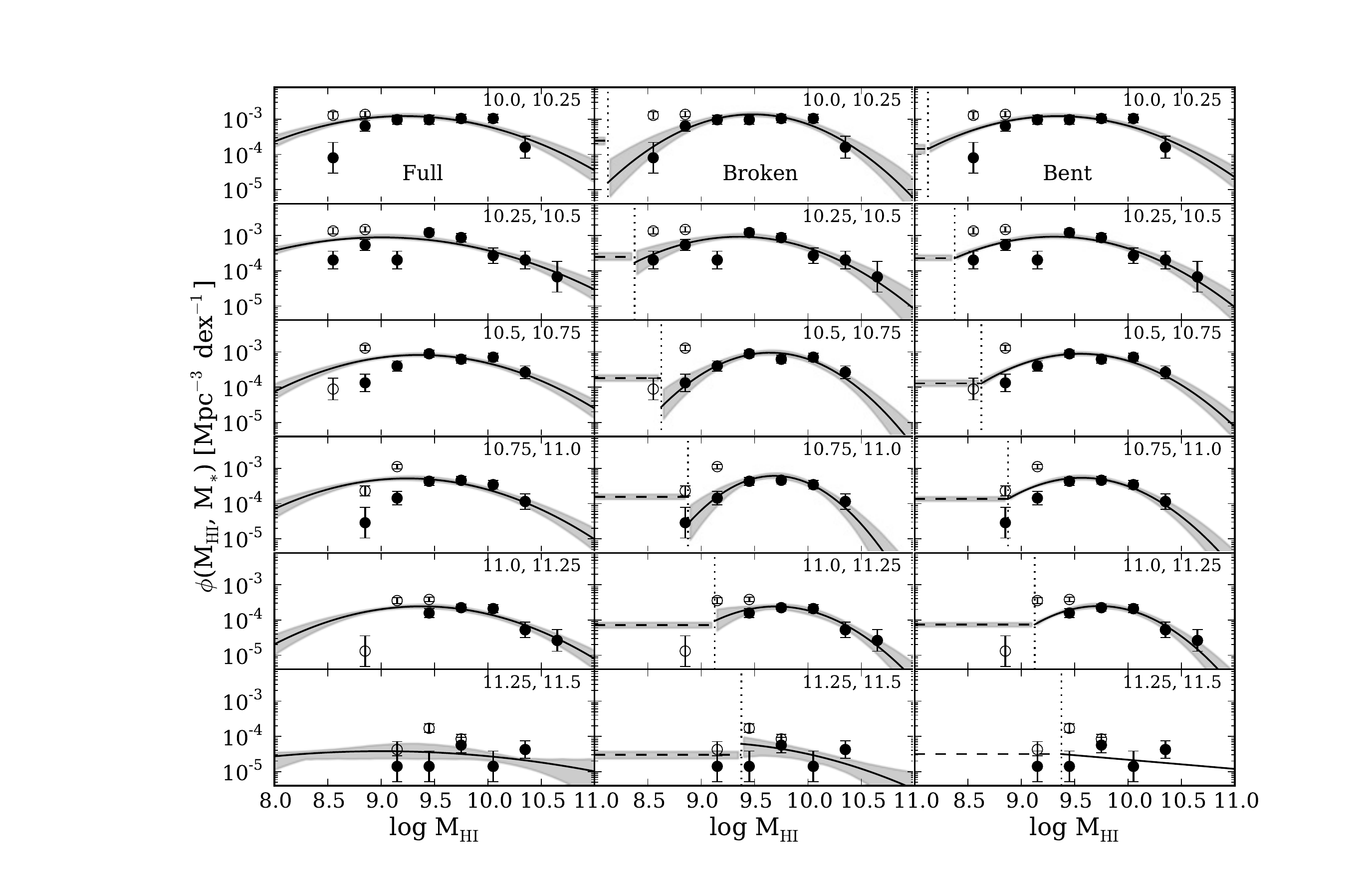}
\caption{Same as Fig. \ref{fig:schechter} but showing three variations of the log-normal function fit to the HISMF.} 
\label{fig:lognormal} 
\end{figure*}

\begin{deluxetable}{cccc}  
\tabletypesize{\scriptsize}
\tablecolumns{4}
\tablecaption{Binned Data for Bivariate HISMF}
\tablewidth{0pt}
\tablehead{
\colhead{log M$_{*}$} & \colhead{log MHI} & \colhead{log $\phi(M_{HI}, M_{*})_{all}$} & 
\colhead{log $\phi(M_{HI}, M_{*})_{det}$}  \\
\colhead{(1)} & \colhead{(2)} & \colhead{(3)} & \colhead{(4)}
}
\startdata
10.0, 10.25    &8.15	&...	&...\\
10.0, 10.25	&8.45	&-3.49$ \pm$ 0.22	&...\\
10.0, 10.25	&8.75	&-2.72 $\pm$ 0.09	&-3.40 $\pm$ 0.19\\
10.0, 10.25	&9.05	&-2.98 $\pm$ 0.12	&-3.02 $\pm$ 0.13\\
10.0, 10.25	&9.35	&-3.02 $\pm$ 0.13	&-3.02 $\pm$ 0.13\\
10.0, 10.25	&9.65	&-2.95 $\pm$ 0.12	&-2.95 $\pm$ 0.12\\
10.0, 10.25	&9.95	&-3.10 $\pm$ 0.14	&-3.10 $\pm$ 0.14\\
10.0, 10.25	&10.25	&-3.19 $\pm$ 0.15	&-3.19 $\pm$ 0.15\\
10.0, 10.25	&10.55	&...	&...\\
10.0, 10.25	&10.85	&...	&...

\enddata
\tablecomments{This table is available in its entirety in a machine-readable form in the online journal.  A portion is 
shown here for guidance regarding its form and content.  Errors reported are Poisson errors.}
\label{tbl:binned_data}
\end{deluxetable}

\begin{deluxetable}{ccccc}  
\tabletypesize{\scriptsize}
\tablecolumns{5}
\tablecaption{Number of Galaxies in Samples}
\tablewidth{0pt}
\tablehead{
\colhead{} & \multicolumn{2}{c}{GASS} & \multicolumn{2}{c}{ALFALFA} \\
\tableline \\
\colhead{log M$_{*}$} & \colhead{Total}  & \colhead{Detections}  & \colhead{Total}  & \colhead{Detections}  \\
\colhead{(1)} & \colhead{(2)} & \colhead{(3)} & \colhead{(4)} & \colhead{(5)}
}
\startdata
10.0, 10.25	&85	&24	&1351	&308	\\
10.25, 10.5	&84	&31	&1269	&297	\\
10.5, 10.75	&96	&28	&1052	&272	\\
10.75, 11.0	&94	&41	&587	&144	\\
11.0, 11.25	&96	&45	&244	&71	\\
11.25, 11.5	&25	&15	&42	&10	\\
\\
Total	&480	&167	&4545	&1102	
\enddata
\label{tbl:numbers}
\end{deluxetable}

\begin{deluxetable*}{ccccc}  
\tabletypesize{\scriptsize}
\tablecolumns{5}
\tablecaption{Schechter Function Fits to Bivariate HIMF\tablenotemark{a}}
\tablewidth{0pt}
\tablehead{
\colhead{log M$_{*}$}  & \colhead{log M$_{HI}^{*}$} & \colhead{$\alpha$} & \colhead{f} & \colhead{log $\phi^{*}$}  \\
\colhead{[M$_{\odot}$]} & \colhead{[M$_{\odot}$]}  & \colhead{} & \colhead{} & \colhead{[Mpc$^{-3}$ dex$^{-1}$]}  \\
\colhead{(1)} & \colhead{(2)} & \colhead{(3)} & \colhead{(4)} & \colhead{(5)}\\
\tableline
\multicolumn{5}{c}{Full Schechter}
}
\startdata
10.0, 10.25    &9.83 $\pm$ 0.10	&0.55 $\pm$ 0.10	&...........	&-2.53 \\
10.25, 10.5	&9.86 $\pm$ 0.54	&0.40 $\pm$ 0.12	&...........	&-2.73 \\
10.5, 10.75	&9.87 $\pm$ 0.10	&0.64 $\pm$ 0.12	&...........	&-2.68 \\
10.75, 11.0	&9.82 $\pm$ 0.11	&0.57 $\pm$ 0.13	&...........	&-2.91 \\
11.0, 11.25	&10.02 $\pm$ 0.12	&0.49 $\pm$ 0.13	&...........	&-3.30 \\
11.25, 11.5	&10.09 $\pm$ 0.35	&0.34 $\pm$ 0.33	&...........	&-3.97 \\
\cutinhead{Broken Schechter}
10.0, 10.25	&9.73 $\pm$ 0.14	&0.83 $\pm$ 0.29	&0.81 $\pm$ 0.06	&-2.45 \\
10.25, 10.5	&9.78 $\pm$ 0.17	&0.59 $\pm$ 0.30	&0.71 $\pm$ 0.07	&-2.66 \\
10.5, 10.75	&9.62 $\pm$ 0.12	&1.44 $\pm$ 0.41	&0.74 $\pm$ 0.05	&-2.60 \\
10.75, 11.0	&9.51 $\pm$ 0.14	&1.76 $\pm$ 0.62	&0.61 $\pm$ 0.06	&-2.89 \\
11.0, 11.25	&9.94 $\pm$ 0.20	&0.69 $\pm$ 0.54	&0.62 $\pm$ 0.07	&-3.25 \\
11.25, 11.5	&10.54 $\pm$ 10.63	&-0.60 $\pm$ 4.68	&0.49 $\pm$ 0.15	&-4.68 \\
\cutinhead{Bent Schechter}
10.0, 10.25	&9.76 $\pm$ 0.11	&0.78 $\pm$ 0.13	&...........	&-2.46 \\
10.25, 10.5	&9.74 $\pm$ 0.12	&0.70 $\pm$ 0.15	&...........	&-2.62 \\
10.5, 10.75	&9.70 $\pm$ 0.10	&1.13 $\pm$ 0.18	&...........	&-2.59 \\
10.75, 11.0	&9.62 $\pm$ 0.11	&1.26 $\pm$ 0.25	&...........	&-2.83 \\
11.0, 11.25	&9.81 $\pm$ 0.14	&1.15 $\pm$ 0.30	&...........	&-3.21 \\
11.25, 11.5	&10.08 $\pm$ 4.40	&0.59 $\pm$ 1.09	&...........	&-3.99 \\

\enddata

\tablenotetext{a}{The fits to the M$_{HI}$ distributions are in the form of a Schechter function such that $\phi(M_{HI}) \ dM_{HI} = \phi^{*} \ (\frac{M_{HI}}{M_{HI}^{*}})^{\alpha} \ e^{\frac{-M_{HI}}{M_{HI}^{*}}} \ dM_{HI}$.  Reported errors for the Schechter parameters $\alpha$, M$_{HI}^*$, and $f$ are 1$\sigma$ uncertainties determined from the MCMC parameters.  Values for $\phi^*$ are based on the median Schechter parameters.}
\label{tbl:schechter}
\end{deluxetable*}

\begin{deluxetable*}{ccccc}  
\tabletypesize{\scriptsize}
\tablecolumns{5}
\tablecaption{Log-normal Fits to Bivariate HIMF\tablenotemark{a}}
\tablewidth{0pt}
\tablehead{
\colhead{log M$_{*}$}  & \colhead{$\mu$} & \colhead{$\sigma$} & \colhead{f} & \colhead{log $\phi^{*}$}  \\
\colhead{[M$_{\odot}$]} & \colhead{[M$_{\odot}$]}  & \colhead{} & \colhead{} & \colhead{[Mpc$^{-3}$ dex$^{-1}$]}  \\
\colhead{(1)} & \colhead{(2)} & \colhead{(3)} & \colhead{(4)} & \colhead{(5)}\\
\tableline
\multicolumn{5}{c}{Full Log-normal}
}
\startdata

10.0, 10.25    &9.22 $\pm$ 0.08	&0.67 $\pm$ 0.08	&...........	&-2.69 \\
10.25, 10.5	&9.01 $\pm$ 0.09	&0.76 $\pm$ 0.10	&...........	&-2.77 \\
10.5, 10.75	&9.36 $\pm$ 0.08	&0.62 $\pm$ 0.07	&...........	&-2.89 \\
10.75, 11.0	&9.24 $\pm$ 0.08	&0.62 $\pm$ 0.08	&...........	&-3.09 \\
11.0, 11.25	&9.38 $\pm$ 0.08	&0.62 $\pm$ 0.08	&...........	&-3.42 \\
11.25, 11.5	&9.01 $\pm$ 1.57	&1.23 $\pm$ 9.35	&...........	&-3.93 \\
\cutinhead{Broken Log-normal}
10.0, 10.25	&9.48 $\pm$ 0.08	&0.46 $\pm$ 0.07	&0.76 $\pm$ 0.05	&-2.81 \\
10.25, 10.5	&9.36 $\pm$ 0.15	&0.53 $\pm$ 0.12	&0.71 $\pm$ 0.07	&-2.91 \\
10.5, 10.75	&9.65 $\pm$ 0.15	&0.38 $\pm$ 6.44	&0.72 $\pm$ 0.06	&-3.04 \\
10.75, 11.0	&9.69 $\pm$ 0.07	&0.32 $\pm$ 0.06	&0.61 $\pm$ 0.06	&-3.30 \\
11.0, 11.25	&9.70 $\pm$ 0.23	&0.42 $\pm$ 0.12	&0.62 $\pm$ 0.07	&-3.60 \\
11.25, 11.5	&9.00 $\pm$ 0.53	&0.81 $\pm$ 3.71	&0.43 $\pm$ 0.12	&-3.86 \\
\cutinhead{Bent Log-normal}
10.0, 10.25	&9.34 $\pm$ 0.06	&0.59 $\pm$ 0.05	&...........	&-2.75 \\
10.25, 10.5	&9.31 $\pm$ 0.06	&0.56 $\pm$ 0.06	&...........	&-2.89 \\
10.5, 10.75	&9.55 $\pm$ 0.05	&0.47 $\pm$ 0.04	&...........	&-2.98 \\
10.75, 11.0	&9.56 $\pm$ 0.04	&0.41 $\pm$ 0.04	&...........	&-3.25 \\
11.0, 11.25	&9.73 $\pm$ 0.04	&0.39 $\pm$ 0.04	&...........	&-3.62 \\
11.25, 11.5	&...............    &...............    &...........    &..... \\

\enddata
\tablenotetext{a}{The fits to the M$_{HI}$ distributions are log-normal such that $\phi({M_{HI}, M_{*}})$ = a $\times$ exp($\frac{-(log M_{HI}-\mu)^{2}}    {2 \sigma ^{2}}$).}
\label{tbl:logn}
\end{deluxetable*}

In this section we present the results of the six models described above.  We attempt to make some determination of which is the better fit, though we note that none of these simple parametrizations can precisely capture the true distribution of HI masses.  In Fig. \ref{fig:schechter} we show $\phi(M_{HI}, M_{*})$, the binned HISMF for three variations of the Schechter function, along with the data used in the fits.  To display the data, we binned the detections and upper limits into HI mass bins 0.3 dex wide.  In each panel of Fig. \ref{fig:schechter}, detections are shown as black circles and the full sample including upper limits for HI non-detections is indicated by empty circles.   As stellar mass increases, the open circles identifying upper limits appear at progressively higher HI masses because of the GASS detection threshold, which is based on HI fraction.  Error bars indicate Poisson uncertainties.  The reported volume densities in each stellar mass bin have been corrected according to the effective volume of the GASS sample in each bin, which is the number of GASS galaxies in each bin divided by the expected space density of galaxies in that bin according to the \citet{Borch2006} stellar mass function.  The binned data are given in Table \ref{tbl:binned_data} and the numbers of galaxies in each stellar mass bin are shown in Table \ref{tbl:numbers}.  We also show a histogram of the detections and upper limits in each HI mass-stellar mass bin in Fig. \ref{fig:nplot}.  We present the binned data only to describe the sample and compare the data to the fits derived from the data.  We emphasize that the fits are not derived from binned HI measurements; rather, each individual HI detection and upper limit contributes to the total probability of the fit independently.

Each column in Fig. \ref{fig:schechter} represents one of the three Schechter functions: full on the left, broken in the middle, and bent on the right.  Each row represents a stellar mass range from log M$_{*}$ = 10.0 to log M$_{*}$ = 11.5; the stellar mass interval is indicated in the upper right corner of each panel.  The best-fit parameters and their 1$\sigma$ uncertainties determined from the MCMC fit are listed in Table \ref{tbl:schechter}.  1$\sigma$ uncertainties on the fit for each HI mass bin are also reflected in the shaded regions surrounding each curve and were determined by computing $\phi$(M$_{HI}$, M$_*$) using all 5000 \emph{emcee} iterations. 

First we examine qualitatively how the HISMF varies with stellar mass and from model to model.  Overall there is remarkable similarity between the HISMF at different stellar masses and for different parametrizations of the Schechter function.  The main difference with respect to stellar mass is that the  HISMF reflects the stellar mass function, which decreases with stellar mass.  Because fewer galaxies are used to fit the models at high stellar masses and the fraction of non-detections increases, the error bars generally widen with stellar mass.  Though the shape of the functions is close to invariant with respect to stellar mass, galaxies in the range 10.5 $<$ log M$_{*}$/M$_{\odot}$ $<$ 11.0 tend to have higher values of $\alpha$, which makes the slope of the biariate HIMF for moderate HI masses steeper, and slightly smaller values for M$_{HI}^*$, so the function cuts off at lower HI masses.  

We also find that the shape of the HISMF is largely independent of the parametrization used.  Though the two pieces of the broken Schechter function are not required to match as they are in the bent function, they almost do, such that the broken Schechter function closely resembles the bent Schechter function.  In this case, the constraint on the bent Schechter function does not seem to bias the fit.  (We will show an example later in which it does.)  Moreover, the calculated space density for galaxies below M$_{HI,break}$ is very similar for the broken and bent cases, showing that the two parametrizations consistently identify a similar fraction of galaxies with a gas fraction above 1\%.  One of the main differences among the three parametrizations is that the error bars for the broken function in the region described by $\alpha$ are significantly larger.  It is reasonable for the error bars to be larger here because there are few detections in this narrow range of HI masses contributing to the estimate of $\alpha$.  Thus the small error bars for the bent Schechter functions perhaps indicate that the constraint that the entire function be continuous does affect $\alpha$ for the bent Schechter function.

Finally, in Fig. \ref{fig:lognormal} we present the results of the three log-normal fits to the data.  The best-fit parameters and associated uncertainties are listed in Table \ref{tbl:logn}.  Because the log-normal function is symmetric, it predicts a much higher space density of galaxies at high HI masses than does the Schechter function.  The three variations of the log-normal functions are similar to the three Schechter functions in that the fits are largely independent of stellar mass except for what was noted above for galaxies with intermediate stellar masses.

A closer look at the fits for the highest stellar mass bin (11.25 $<$ log M$_*$/M$_{\odot}$ $<$ 11.5) reveals that they do not exhibit the same shape as do the fits at lower stellar masses.  Moreover, the size of the error regions indicates that these fits are not well constrained (see also reported errors in Tables \ref{tbl:schechter} and \ref{tbl:logn} and plots below).  These results can be attributed to the small number of galaxies in this stellar mass bin, a number that is less than a third of the number of galaxies in the other stellar mass bins (Table \ref{tbl:numbers}) despite attempts to maintain a flat stellar mass distribution for the GASS sample.  We choose to keep this stellar mass bin in our analysis to show where our ability to accurately describe the shape of the HISMF breaks down.  More observations at higher stellar masses (and the final GASS data release, Catinella et al., MNRAS submitted) will be necessary to understand how the HISMF changes for the most massive galaxies in the local universe.

\begin{figure}[t]
\epsscale{1.2}
\plotone{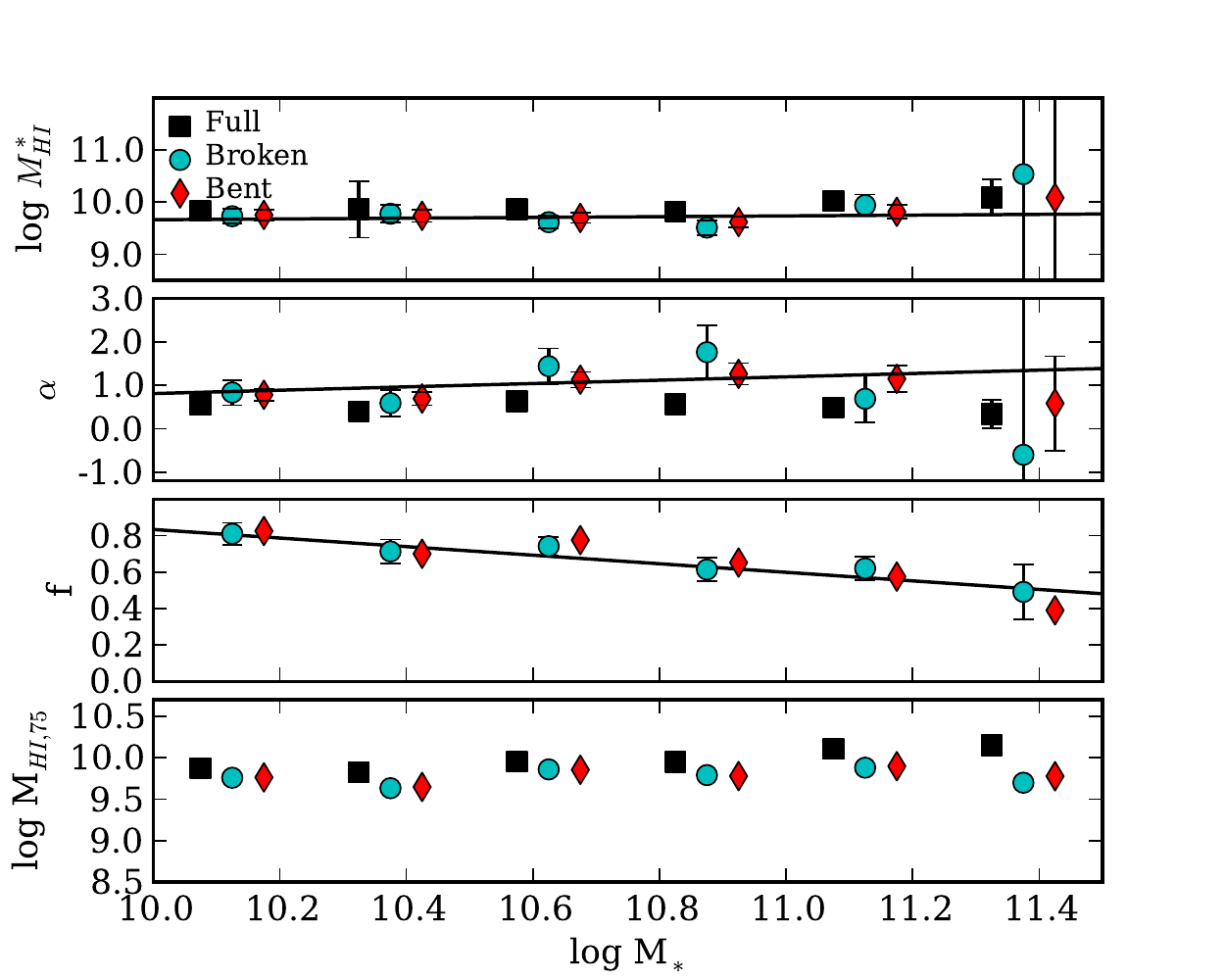}
\caption{Schechter parameters $\alpha$, log M$_{HI}^*$, $f$, and M$_{HI,75}$ vs. stellar mass and the best-fit lines as a function of stellar mass from Section \ref{sec:otherfits}.} 
\label{fig:params} 
\end{figure}

\begin{figure}[t]
\epsscale{1.2}
\plotone{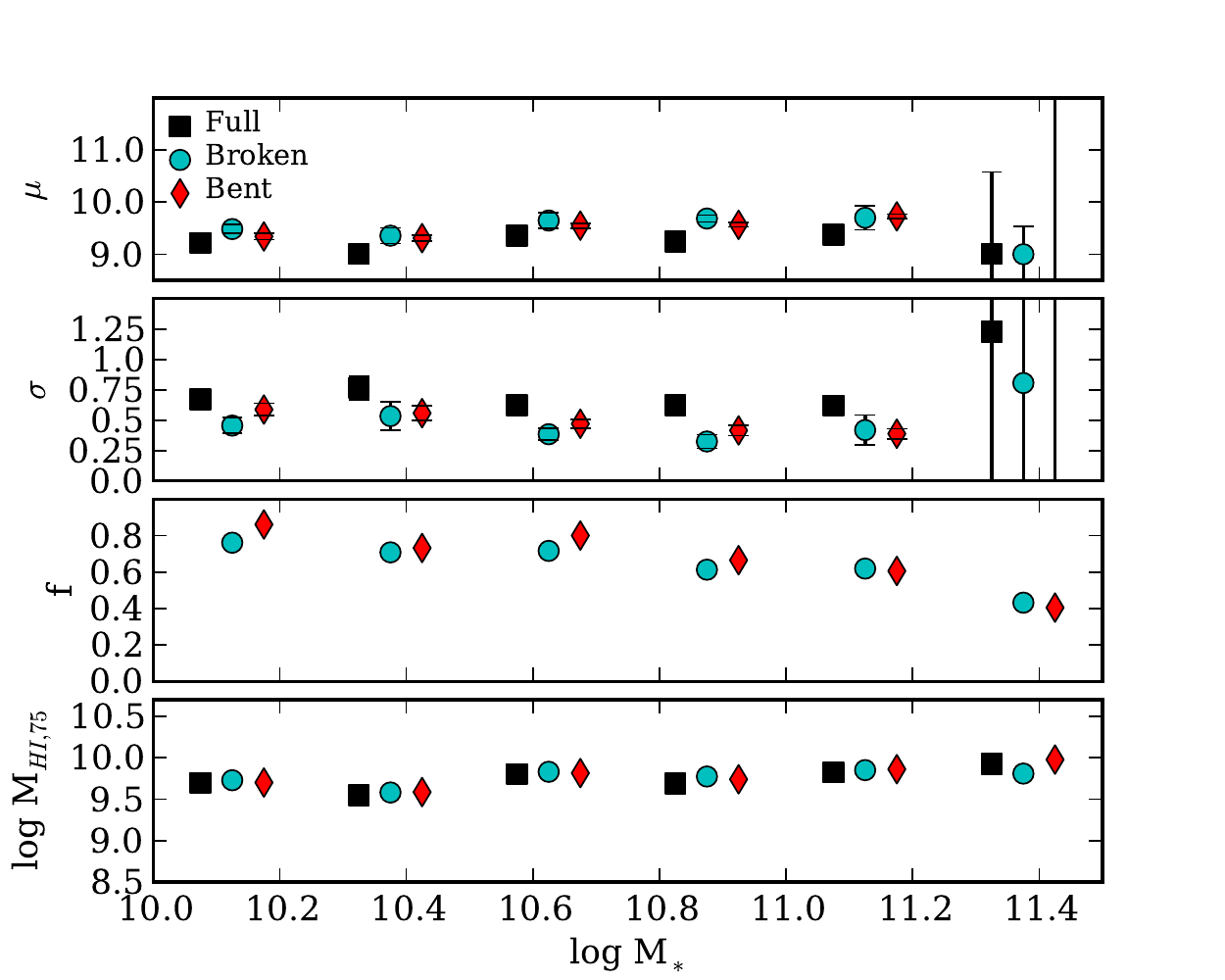}
\caption{Log-normal parameters $\sigma$, $\mu$, $f$, and M$_{HI,75}$ vs. stellar mass.} 
\label{fig:params2} 
\end{figure}
 
In Figs. \ref{fig:params} and \ref{fig:params2} we quantitatively examine how the best-fit Schechter and log-normal parameters vary with stellar mass.  The top three panels show log M$_{HI}^*$, $\alpha$ and $f$ vs. stellar mass for the three parametrizations of the Schechter function.  This plot quantitatively confirms the trends we discussed above.  The full, broken and bent fits yield values for M$_{HI}^*$ that are close to constant with respect to stellar mass.  We will discuss the relative constancy of M$_{HI}^*$ in Section \ref{sec:constantmstar}.  The broken and bent Schechter models produce $\alpha$'s that range between 0.5 and 1.2 with no obvious trend with stellar mass.  The full Schechter function yields values for $\alpha$ that are slightly lower than the other $\alpha$'s ($\sim$ 0.5).
The third panel shows $f$ vs. stellar mass for the broken and bent models.  This value is a parameter in the broken model and can be derived from the results of the bent model.  The effective $f$ for the bent fit and $f$ for the broken fit track each other nicely, confirming that the fits yield similar HISMFs.  (As an additional check they are also comparable to the observed detection fractions reported in \citet{Catinella2010, Catinella2012}.)  Both models show that the fraction of galaxies with gas fractions above 1\% decreases with stellar mass.  As another way of comparing the three Schechter functions, in the bottom panel we show M$_{HI,75}$ for each of the three fits, where M$_{HI,75}$ is the HI mass in each stellar mass bin below which lie 75\% of the galaxies in that stellar mass bin.  The broken and bent Schechter functions yield values for M$_{HI,75}$ that are nearly identical.  The values for the full Schechter function are similar to those of the truncated Schechter functions but diverge slightly from them at high stellar masses.

Fig. \ref{fig:params2} shows a similar analysis of the parameters of the log-normal fits.  The trends for M$_{HI,75}$ and $f$ vs. stellar mass are similar to those for the Schechter functions.  $\mu$, which is related to M$_{HI}^*$ for the Schechter fit, does not show an obvious correlation with stellar mass.  $\sigma$, which measures the width of the distribution, decreases slightly with stellar mass from about 0.6 to 0.4.

\subsection{Covariance of Parameters}

\begin{figure*}[t]
\epsscale{0.8}
\plotone{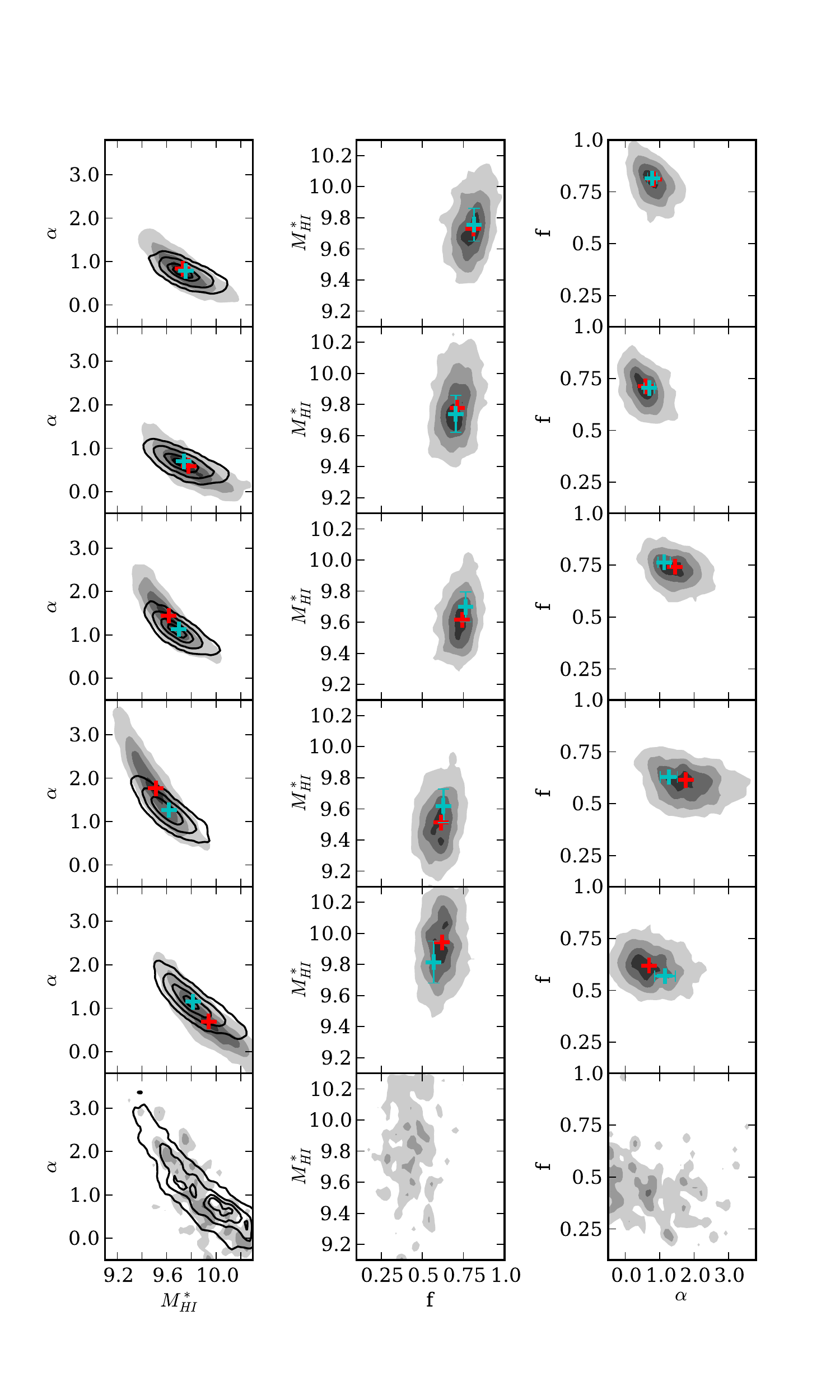}
\caption{Error contours for Schechter parameters $\alpha$, M$_{HI}^*$ and $f$ for the broken (grayscale contours; red crosses) and bent (black contours; cyan crosses) models in six stellar mass bins.  Contours enclose 10, 25, 50, and 95\% of the 5000 Schechter parameters representing the 5000 MCMC iterations.} 
\label{fig:contours} 
\end{figure*}

Next we wish to examine the relationship between the model parameters to uncover any covariance.  As an example, we do this for the broken and bent Schechter fits in Fig. \ref{fig:contours}.  Each row shows a different stellar mass bin corresponding to the six rows in Fig. \ref{fig:schechter}.  In each panel we show the error contours associated with two parameters for both functions.  The contours enclose 10, 25, 50, and 95\% of the 5000 Schechter parameters ($\alpha$, M$_{HI}^*$, $f$) representing the 5000 MCMC iterations for each stellar mass bin.

In the left column the tilt in the contours for $\alpha$ vs. M$_{HI}^*$ shows that the two Schechter parameters for the broken and bent Schechter functions exhibit some covariance along the direction in which they are anti-correlated.  A high $\alpha$ and a low M$_{HI}^*$ could describe the distribution just as well as a low $\alpha$ and a high M$_{HI}^*$.  We comment on the origin of this covariance in Section \ref{sec:otherfits}.  The narrower contours for the bent fit show that $\alpha$ is more tightly constrained by the model, likely because of the requirement that the Schechter function smoothly connect to the flat part of the function at M$_{HI,break}$.  Though there are some noticeable differences between the two fits at high stellar masses, the contours for the two fits generally overlap across the entire stellar mass range, indicating that any difference between the two model fits is small compared to the errors in the parameters.

The middle column shows the relationship between M$_{HI}^*$ and $f$ and the right column shows $f$ vs. $\alpha$.  For the bent fit we only show the effective $f$.  We find that the broken fit yields a tight constraint on $f$, which is not surprising since the GASS survey is designed to provide an accurate measure of this quantity.  We also find that there is little obvious covariance between f and $\alpha$.

\subsection{Model fit at high HI masses and comparison with ALFALFA}
\label{sec:highmass}

An important distinction between the two classes of models considered above (Schechter vs. log-normal) is the shape of the function at high HI masses.  Here we explore two questions: 1) whether GASS includes sufficient numbers of HI-rich galaxies to provide a strong constraint on the function at the HI-rich end and 2) whether we can favor one class of model over the other based on the success of our fits.  To test these questions we first conduct an analysis that makes use of an enlarged sample of HI-rich galaxies obtained using the ALFALFA survey.  We then use the sum of each model across stellar masses to provide sufficient signal to allow us to distinguish between model classes.

GASS was designed to sensitively detect HI in massive HI-poor galaxies but does not detect as many HI-rich galaxies as do other shallower blind surveys with significant sky coverage, such as ALFALFA \citep{Giovanelli2005}.  When complete, ALFALFA will have scanned over 7000 square degrees of the sky, so the number of ALFALFA detections in the GASS stellar mass range exceeds the number of GASS detections.  We refine the measurement of the HISMF at high HI masses by including in the fits detections and upper limits from the ALFALFA survey.   

The ALFALFA sample we use includes all galaxies in the GASS parent sample that lie in the regions of the sky already observed and cataloged by ALFALFA as of the $\alpha$.40 release \citep{Haynes2011}.  In this region 1102 galaxies were detected by ALFALFA and 3443 were not detected (see Table \ref{tbl:numbers}).  We calculate upper limits for the non-detections by using the ALFALFA integrated flux limit, S$_{lim}$ in Jy km s$^{-1}$, from \citet{Martin2010}.  S$_{lim}$ depends on the signal-to-noise ratio (S/N) and velocity width (W$_{50}$) of the HI line:

\begin{equation}
S_{lim} = \left\{ \begin{array}{ll}
         0.15 \ S/N \ (W_{50}/200)^{1/2} & \mbox{if $W_{50} < 200 km s^{-1}$};\\
        0.15 \ S/N \ (W_{50}/200)& \mbox{if $W_{50} \ge 200 km s^{-1}$}.\end{array} \right. 
\end{equation}

We let S/N = 5.0 and W$_{50}$ = 300 km $s^{-1}$ to be consistent with the calculation of upper limits for GASS.  \citet{Martin2010} calculate HI masses according to:

\begin{equation}
M_{HI} = 2.356 \times 10^{5} D^{2}_{Mpc} \ S_{lim}.
\end{equation}

Thus, the upper limits we calculate for ALFALFA non-detections are simply a function of distance:

\begin{equation}
M_{HI_{UL}} = 2.65 \times 10^{5} D^{2}_{Mpc}
\end{equation}

To derive the HISMF that accounts for the GASS and ALFALFA surveys together we simply add the log likelihoods for the galaxies in each sample.  (We checked with our simulation that applying different weights to each sample does not significantly affect the result.)  As we mentioned above, the log likelihoods include the error on each HI measurement.  We derive errors for the ALFALFA detections in the same way we did for GASS.  We also set the error for ALFALFA upper limits to 0.3 dex.

\begin{figure}[t]
\epsscale{1.0}
\plotone{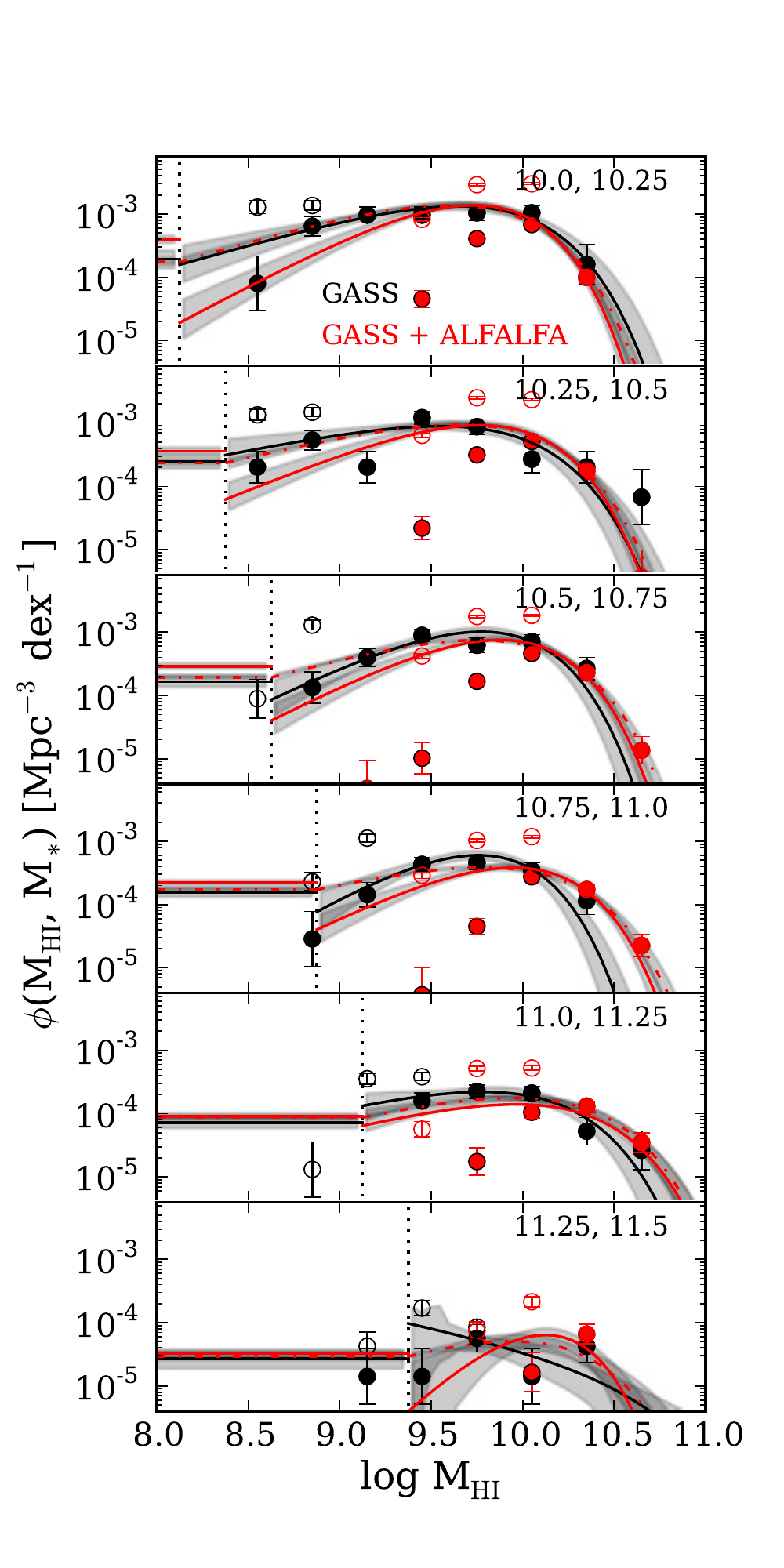}
\caption{The broken Schechter fit to the GASS HISMF (black solid line) compared to the broken (solid red) and bent (dotted red) Schechter fits to the HISMF that includes ALFALFA detections and upper limits.  Black circles are as in Fig. \ref{fig:schechter}.  Solid red circles show ALFALFA detections; open red circles show ALFALFA upper limits.} 
\label{fig:alfalfa} 
\end{figure}

In Fig. \ref{fig:alfalfa} we present the ALFALFA data and the results of the joint fit to GASS and ALFALFA.  We show the ALFALFA data in the same way we show the GASS data: detections are shown as solid symbols and the full sample including upper limits is shown as open symbols.  The ALFALFA data generally extend to higher HI masses (log M$_{HI}$/M$_{\odot}$ $\ge$ 10.5) than the GASS data, which end at log M$_{HI}$/M$_{\odot}$ $\sim$ 10.25 in most stellar mass bins.  Though ALFALFA provides better coverage of the HI-rich end of the HIMF, the small number of galaxies detected by ALFALFA below log M$_{HI}$/M$_{\odot}$ $\sim$ 10.0 emphasizes the need for deeper surveys such as GASS that can provide HI measurements of massive galaxies with smaller amounts of HI.

Fig. \ref{fig:alfalfa} compares the GASS broken Schechter fit to the bent and broken versions of the GASS + ALFALFA fit.  First we will examine the HI-rich end of the HISMF. In general both variations of the GASS+ALFALFA fits predict slightly fewer HI-rich galaxies at low stellar masses and slightly more HI-rich galaxies at high stellar masses.  The major difference between the two fits is in the range 10.75 $<$ log M$_{*}$/M$_{\odot}$ $<$ 11.0, where the GASS+ALFALFA fit extends to higher HI masses.  We noted previously that the GASS fits in this mass range predicted surprisingly low space densities.  The GASS+ALFALFA fits are more consistent with respect to stellar mass, perhaps making up for small anomalies in the GASS sample.  Though the fits in this mass range argue that ALFALFA data are necessary to properly describe the HI-rich end of the HISMF, we see this mass range as an exception rather than the rule.  We will show later that a continuous fit applied to only GASS data is able to account for the anomaly in the stellar mass bin 10.75 $<$ log M$_{*}$/M$_{\odot}$ $<$ 11.0 without including ALFALFA galaxies.  Thus, the binned fit, rather than the use of only GASS data, is the cause of the discrepancy in this stellar mass bin.

At lower HI masses, the broken and bent GASS+ALFALFA fits vary significantly.  The broken Schechter fit exhibits a steep slope down to M$_{HI,break}$ while the constraints of the bent Schechter fit cause the slope to be much shallower.  The significant difference between $\alpha$ for the bent and broken fits with ALFALFA contrasts with the relative similarity between $\alpha$ for the bent and broken fits to GASS data only.  It is likely that the lack of ALFALFA data at low HI masses had too much influence on the fit here, biasing the broken fit to predict low numbers of galaxies within this HI mass range.  For ease of analysis and because we have shown the GASS sample to provide a sufficient estimate of the HISMF at high HI masses, we do not include ALFALFA in each of our subsequent fits.  Future work may continue to explore the combined use of GASS and ALFALFA.

\begin{figure*}[t]
\epsscale{1.0}
\plotone{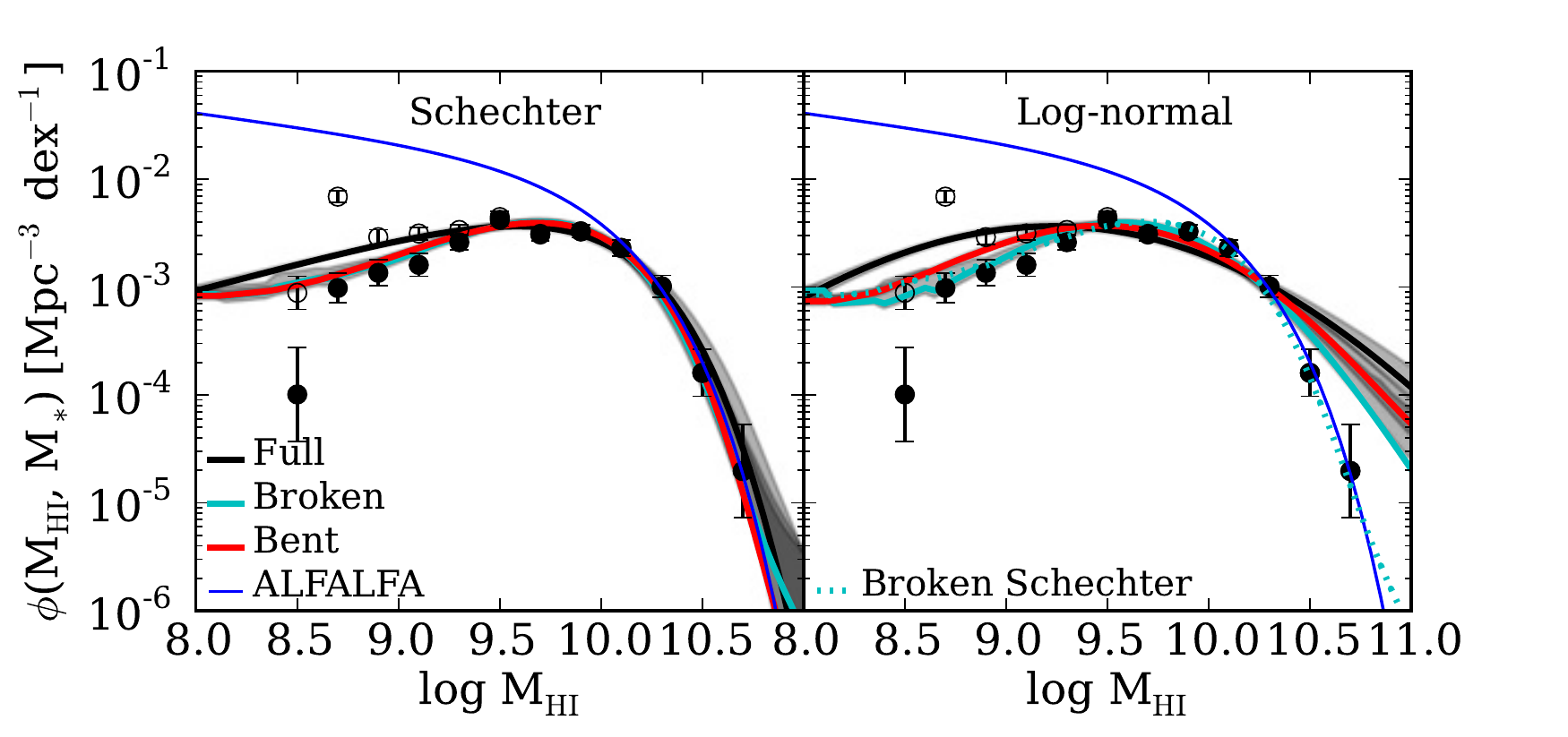}
\caption{The total HISMF for massive galaxies based on six different models compared to the ALFALFA HIMF \citep[blue][]{Martin2010}.  \emph{Left}:  The sum of the full, broken, and bent Schechter fits to the HISMF.  \emph{Right}: The sum of the full, broken, and bent log-normal fits to the HISMF.} 
\label{fig:total} 
\end{figure*}

\subsection{Model Selection and Validation}
\label{sec:modelselection}

We have produced six fits to the HISMF based on six different models and we wish to select which is the best match to the data.  To distinguish between the models quantitatively we calculated several parameters that are commonly used to assess goodness-of-fit or for model selection.  We list these numbers in the Appendix and briefly discuss them here.  They include the probabilities calculated in the MCMC runs, the Bayesian Information Criterion \citep[BIC;][]{Schwarz1978}, the Poisson probabilities, and the $\chi^2$ statistic.

In Table \ref{tbl:prob} we list the MCMC probabilities and Bayesian Information Criterion for each model and each stellar mass bin.  Both measures of the probabilities are in the same range for each stellar mass bin and do not strongly favor one model over the others.

We can examine the qualitative differences among the six models in two ways: by comparing the integrated model HISMFs, which emphasize deviations from the data, and by considering the limitations of each model that are not quantifiable.  We derive the total HISMF for massive galaxies (Fig. \ref{fig:total}) by summing the HISMF across all stellar mass bins for each model.  Because M$_{HI,break}$ varies with stellar mass and the total HISMF is a sum over all stellar mass bins, the total HISMFs for the broken and bent fits do not have the same characteristic flat regions at low HI masses.  At the lowest HI masses the shape of the HISMF is still poorly constrained by our data.

First we examine the differences between the Schechter functions and log-normal functions, which are shown separately in the panels on the left and right.  Though the Schechter and log-normal functions predict a similar number of galaxies at low HI masses, each of the three log-normal fits systematically predicts more galaxies at high HI masses than do the three Schechter fits.  The space densities they predict are well above the data points and their error bars.  Based on this discrepancy between the data and the log-normal fits to the data, we can safely conclude that the Schechter function's steeper dropoff at high HI masses makes it a better approximation of the true HISMF 

To test this conclusion we quantitatively compare the predictions of the six models to the data by calculating the $\chi^2$ statistic and the Poisson probability for HI masses above log M$_{HI}$/M$_{\odot}$ = 10.  These numbers (see Table \ref{tbl:probtot}) confirm that the Schechter function is a much better description of the data at high HI masses.

\citet{Solanes1996} conducted a test similar to ours by comparing Schechter and Gaussian fits to the total HIMF derived from 934 giant spirals in an HI flux limited sample.  They found that the Gaussian and Schechter fits describe the full sample and the sample divided by morphology equally well.  They note that discrepancies between the two fits occur at low (log M$_{HI}$/M$_{\odot}$ $<$ 9.0) and high (log M$_{HI}$/M$_{\odot}$ $>$ 10.5) HI masses.  It is in these HI mass ranges where we also find the greatest discrepancy between the two fits.

We cannot favor one version of the Schechter fit over another based on their total HISMFs because they are virtually identical, which Table \ref{tbl:probtot} confirms.  Instead it is important to consider the limitations inherent to each model.  Upper limits can significantly affect the slopes of the full Schechter and log-normal models at moderate HI masses.  The break in the broken and bent models allows us to use upper limits to estimate the space density of galaxies with low HI fractions while not allowing upper limits to dictate the shape of the function.

A limitation of the bent model is that the two sections of the function are required to meet.  This requirement seems reasonable because a true disconnect in a mass function (such as in the broken model) does not make physical sense.  But this requirement is only reasonable if we are confident in the shape we impose on both sides of the break.  A flat distribution at low HI masses is not the only choice and we discuss this in Section \ref{sec:lowHImass}.

Because of the obvious limitations of the full and bent models, we choose to proceed with the broken Schechter function.  We emphasize, though, that this parametrization is not necessarily the best or most precise way of describing the HISMF but reflects our desire to provide a model for the data that accurately describes the true distribution in an intuitive way with as few parameters as needed.   

In Fig. \ref{fig:total} we also compare our total HISMF for massive galaxies to the HIMF from ALFALFA \citep{Martin2010}.  The GASS HISMF matches the ALFALFA curve at high HI masses but misses many of the HI-poor galaxies that are detected in ALFALFA because they are below the GASS stellar mass and redshift cutoffs (e.g. dwarf galaxies).  The agreement between the two functions at high HI masses indicates that many of the most HI-rich galaxies in the local universe are in fact galaxies that are also massive in stars (any contribution to the HI-rich end of the function from lower stellar mass galaxies would be small).  This was also seen and discussed in \citet{Schiminovich2010}.  We return to this point below.

\begin{deluxetable*}{ccccccc}  
\tabletypesize{\scriptsize}
\tablecolumns{7}
\tablecaption{$\Omega_{HI}$}
\tablewidth{0pt}
\tablehead{
\colhead{log M$_{*}$} & \colhead{log $\Omega_{HI,M_*}$}\tablenotemark{a} & \multicolumn{5}{c}{$\Omega_{HI,M_*}$ / $\Omega_{HI}$}\tablenotemark{b} \\
\tableline \\
\colhead{} & \multicolumn{2}{c}{Broken Schechter} & \colhead{Bivariate} & \colhead{Trivariate} & \colhead{Dets.} & \colhead{All}\tablenotemark{c}
}
\startdata
10.0, 10.25       &-4.24	&0.13	&0.12	&0.12	&0.13	&0.14\\
10.25, 10.5	&-4.42	&0.09	&0.11	&0.09	&0.09	&0.09\\
10.5, 10.75	&-4.37	&0.10	&0.09	&0.10	&0.10	&0.10\\
10.75, 11.0	&-4.66	&0.05	&0.06	&0.06	&0.05	&0.06\\
11.0, 11.25	&-4.84	&0.03	&0.03	&0.04	&0.03	&0.04\\
11.25, 11.5	&	&0.01	&0.01	&0.01	&0.01	&0.01\\
\tableline
Total		    &-3.75	&0.41	&0.41	&0.42	&0.41	&0.44\\
\enddata
\tablenotetext{a}{Typical errors are 0.1 assuming a 1$\sigma$ error on the space density.}
\tablenotetext{b}{$\Omega_{HI,M_*}$ as a fraction of $\Omega_{HI}$ reported in \citet{Martin2010}.}
\tablenotetext{c}{The sum of detections and non-detections where we have set the HI mass of non-detections equal to their upper limits.}
\label{tbl:omegaHI}
\end{deluxetable*}

We calculate $\Omega_{HI,M_* > 10^{10}}$, the ratio of the HI density contributed by massive galaxies to the critical density, and present our results for each stellar mass bin in Table \ref{tbl:omegaHI}.  In each bin we compute $\Omega_{HI,M_* > 10^{10}}$ by integrating the broken Schechter fit to the HISMF.  We do the same with the bivariate and trivariate fits (see Section \ref{sec:otherfits}) and we compare the results to a simple sum of the HI detections and upper limits.  The total HI density in the GASS stellar mass range, obtained by summing $\Omega_{HI}$ in each stellar mass bin for the broken Schechter function, is log $\Omega_{HI}$ = -3.75, which is 41\% of the total $\Omega_{HI}$ in the local universe derived from ALFALFA \citep{Martin2010}.  This HI density agrees with that reported in \citet{Schiminovich2010}, which represents 42\% of the \citet{Martin2010} value.  The agreement between their calculation, which is based on a simple sum of the observed HI masses, and ours validates our model of the HISMF.  A sum of the GASS detections and upper limits (see final column) overestimates $\Omega_{HI,M_* > 10^{10}}$ because the upper limits are included as true HI masses in the sum whereas our fitting method assumes the true HI masses can be less than the upper limits.  As \citet{Schiminovich2010} noted, the comparison between $\Omega_{HI}$ derived from ALFALFA with $\Omega_{HI,M_* > 10^{10}}$ derived from GASS emphasizes that massive galaxies contribute significantly to the total HI content in the local universe.  This is not entirely surprising since \citet{Martin2010} show that galaxies with 9.0 $<$ log M$_{HI}$/M$_{\odot}$ $\le$ 10.0 contribute the most to $\Omega_{HI}$ and this is the HI mass range that many of the GASS galaxies occupy.

\section{Discussion}

\subsection{The Shape of the HIMF}
In this paper we explore the shape of the HISMF within six independent stellar mass bins (effectively resulting in 12 or 18 parameter fits) to assess what shapes the distribution of HI masses at different stellar masses.  Later in this section we explore how this shape can be expressed as a function of stellar mass and also SFR, leading to a reduced parametrization, and providing some clues on how the processes that drive the shape of the mass function depend on these physical parameters.

Without a detailed model predicting the shape we adopted a generalized approach, comparing the results of six different fits to the HISMF.  We tested three variations of the Schechter and log-normal functions in order to account for the high fraction of HI non-detections at low HI masses. For two of these variations we imposed a flat distribution on the HI-poor end because we have very little a priori knowledge of the true shape of the distribution.  This form is also easy to interpret as representing the low gas fraction component of the distribution.  We found that the broken Schechter function, in which we fit a Schechter function above a 1\% HI gas fraction and a flat function below, describe the shape of the GASS HISMF better than the log-normal functions and is free of the potential bias of the other Schechter models.

In the following sections we appeal to higher-dimensional fits to the HIMF, simulations that seek to reproduce the HIMF, and other HI observations to understand the physical processes that contribute to the shape of the HISMF.

\subsection{The Dependence of the HI Mass Function on Stellar Mass and Star Formation Rate}
\label{sec:otherfits}

\begin{deluxetable*}{lccccccccc}  
\tabletypesize{\scriptsize}
\tablecolumns{10}
\tablecaption{Continuous Fits\tablenotemark{a}}
\tablewidth{0pt}
\tablehead{
\colhead{} & \colhead{$\alpha_m$} & \colhead{$\alpha_s$} & \colhead{$\alpha_o$} & \colhead{M$_m$} & \colhead{M$_s$} & \colhead{M$_o$} & \colhead{f$_m$} & \colhead{f$_s$} & \colhead{f$_o$}   
}
\startdata
Bivariate Fit	&0.39	&...	&1.00  &0.07	&...	& 9.70 &-0.24	&... &0.72\\
Trivariate Fit	&0.47	&0.95	&1.26	&0.22	&-0.03	&9.63 & -0.20  &0.35  &0.92
\enddata
\tablenotetext{a}{Top row: best-fit parameters for broken Schechter fits where the parameters depend on stellar mass.  Bottom row: best-fit parameters for broken Schechter fits where the parameters depend on stellar mass and SFR.}
\label{tbl:otherfits}
\end{deluxetable*}

\begin{figure*}[t]
\epsscale{1.2}
\plotone{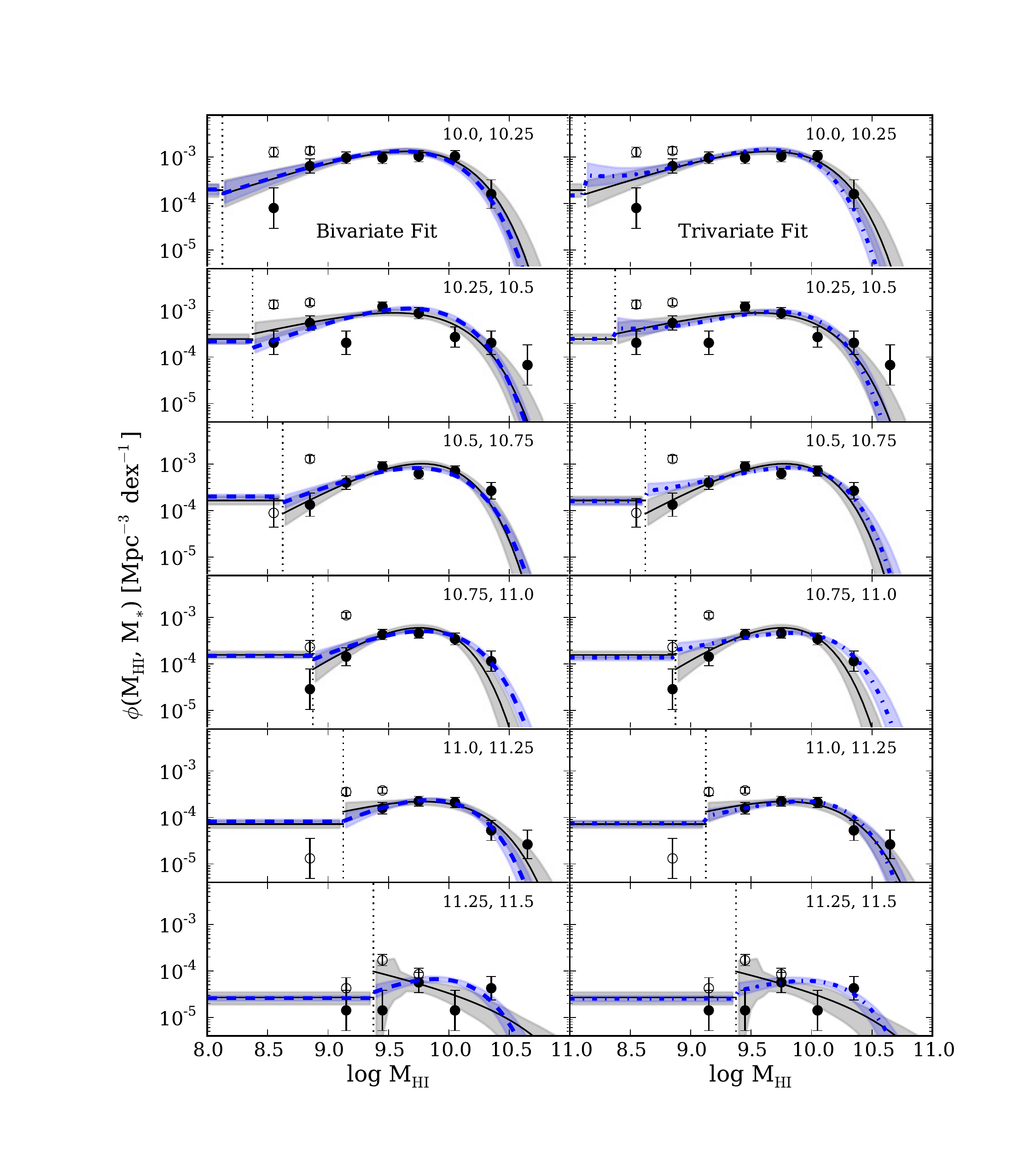}
\caption{The binned broken Schechter HISMF (solid line) compared to the continuous bivariate fit (left; blue dashed line), whose parameters depend on stellar mass, and the continuous trivariate fit (right; blue dotted line), whose parameters depend on stellar mass and SFR.  $\alpha$, M$^*_{HI}$ and $f$ in each bin are calculated as described in the text.} 
\label{fig:stellarmass} 
\end{figure*}

To gain insight into the physical mechanisms that shape the HISMF, we fit two additional models to the data, both of which are continuous fits across the range of stellar masses rather than binned fits within indepedent stellar mass bins.  The first is a continuous bivariate fit whose Schechter parameters ($\alpha$, M$_{HI}^*$, $f$) depend on stellar mass, which yields a variation on the HISMF, $\phi$(M$_{HI}$,M$_*$).  The second is a continuous trivariate fit in which the parameters are functions of stellar mass and SFR.  From this fit we can derive $\phi$(M$_{HI}$, M$_*$, SFR), the HI-M$_*$-SFR function.  As stellar mass is related to the growth history of galaxies and SFR is linked to cold gas content, the variation of the HIMF with respect to these two quantities could provide insight into what shapes the distribution of HI masses at a given stellar mass.  Although galaxies on the star-forming sequence have a well-defined link between SFR and stellar mass \citep[e.g.][]{Brinchmann2004, Salim2007, Schiminovich2007}, at a given stellar mass galaxies exhibit a wide range of SFEs \cite[e.g.]{Schiminovich2010}.  The scatter between SFR and M$_{HI}$, and the additional scatter between those two quantities and stellar mass, suggests that including SFR as a parameter in our model should uncover trends that are otherwise not apparent.

First we assess the results of the continuous bivariate fit, which is shown in the left column of Fig. \ref{fig:stellarmass}.  The best-fit parameters are listed in Table \ref{tbl:otherfits} and the definitions of the parameters are in Table \ref{tbl:otherfits_def}.  The plot was constructed by calculating $\alpha$, M$_{HI}^*$, and $f$ based on the central stellar mass of the previously defined stellar mass bins.  The fits are similar to the original binned fits (with one exception, noted below).  We examine these fits by considering each parameter individually.

The line in Fig. \ref{fig:params} shows how each parameter changes with stellar mass according to this fit.  The results of this higher-dimensional fit track the results of the broken Schechter function fit to the HISMF.  The weak correlation between M$_{HI}^*$ and stellar mass (M$_{HI}^*$ $\propto$ M$_*^{0.07}$) confirms the relative invariance of the fits with respect to stellar mass, which was illustrated in Fig. \ref{fig:schechter}.  The inverse relationship between $f$ and stellar mass reveals that more massive galaxies are less likely to have a high gas fraction, and in particular, one above 1\%.  This result verifies that HI gas fraction is inversely correlated with stellar mass \citep[e.g.][]{Catinella2010}.  

Next we discuss the trivariate fit, whose parameters are defined according to the equations in Table \ref{tbl:otherfits_def}.  This fit reveals the joint effect of stellar mass and SFR on the HIMF and shows that the shape of the HIMF is more strongly dependent on SFR.  The best-fit parameters (see Table \ref{tbl:otherfits}) reveal that $\alpha$ is much more dependent on both stellar mass ($\alpha$ $\propto$ M$_*^{0.47}$) and SFR ($\alpha$ $\propto$ SFR$^{0.95}$) than are M$_{HI}^*$ and $f$.  $\alpha$ is higher ($\sim$2-3) for galaxies with higher stellar masses and higher SFRs, implying that the distribution of HI masses is more sharply peaked.  

In the right column of Fig. \ref{fig:stellarmass} we show how this fit compares to the original fit.  The curve for each stellar mass bin is constructed by dividing the galaxies in that stellar mass bin into bins of SFR and summing the mass functions derived using the central stellar mass of the bin and the central SFR for each SFR-M$_{*}$ bin.  The HIMF for each SFR-M$_*$ bin is weighted by the number of galaxies in it.  This fit generally approximates the original fit at high HI masses.  At low HI masses and low stellar masses, the trivariate fit predicts slightly more galaxies.  This is due to the combination of low stellar masses and low SFRs that produces a low value for $\alpha$ and a mass function that rises towards lower HI masses.  We note that the main difference between the bivariate, trivariate and binned fits is in the stellar mass bin 10.75 $<$ log M$_*$/M$_{\odot}$ $<$ 11.0.  In this stellar mass range the bivariate and trivariate fits predict more HI-rich galaxies and their prediction matches that of the GASS+ALFALFA fit described in Section \ref{sec:highmass}.  These fits appear to do a better job than the binned fits of averaging over anomalies in the GASS data and arriving at the true distribution of HI masses.  Overall these fits seem to match the broken Schechter function fit: $\Omega_{HI,M_* > 10^{10}}$ derived from the bivariate and trivariate fits agree with that derived from the broken Schechter function (see Table \ref{tbl:omegaHI}).  In Table \ref{tbl:probtot} we compare these fits to the binned fits and we find that the continuous fits are generally better than the binned log-normal fits but are no better than the binned Schechter fits.

\begin{figure}[t]
\epsscale{1.3}
\plotone{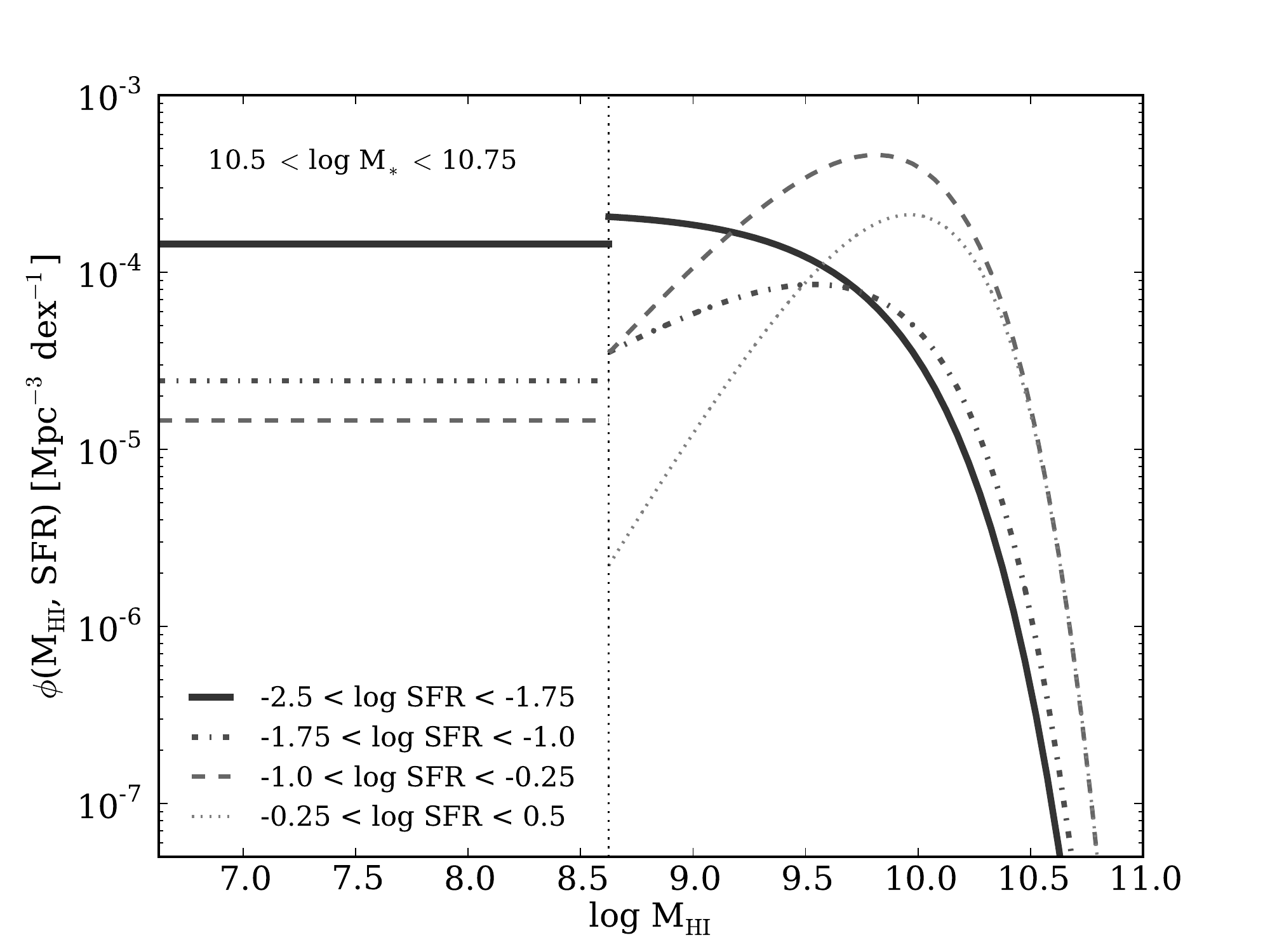}
\caption{An example of the continuous trivariate HI-M$_*$-SFR function for galaxies with stellar masses in the range 10.5 $<$ log M$_*$/M$_{\odot}$ $<$ 10.75 and log SFRs ranging from -2.5 to 0.5.  There is no population of galaxies with M$_{HI}$ $<$ M$_{HI,break}$ in the highest SFR bin in this stellar mass range because $f$ is expected to be close to 1.0.} 
\label{fig:sfrbins} 
\end{figure}

To emphasize the variation in the HI-M$_*$-SFR function with SFR we present in Fig. \ref{fig:sfrbins} the HIMF for galaxies with 10.5 $<$ log M$_*$/M$_{\odot}$ $<$ 10.75 and log SFR ranging from -2.5 to 0.5.  The overall shape of the HIMF, and in particular the slope at the HI-poor end, varies significantly with SFR.  We discuss this in more detail in the following section.

\begin{figure*}[t]
\epsscale{1.0}
\plotone{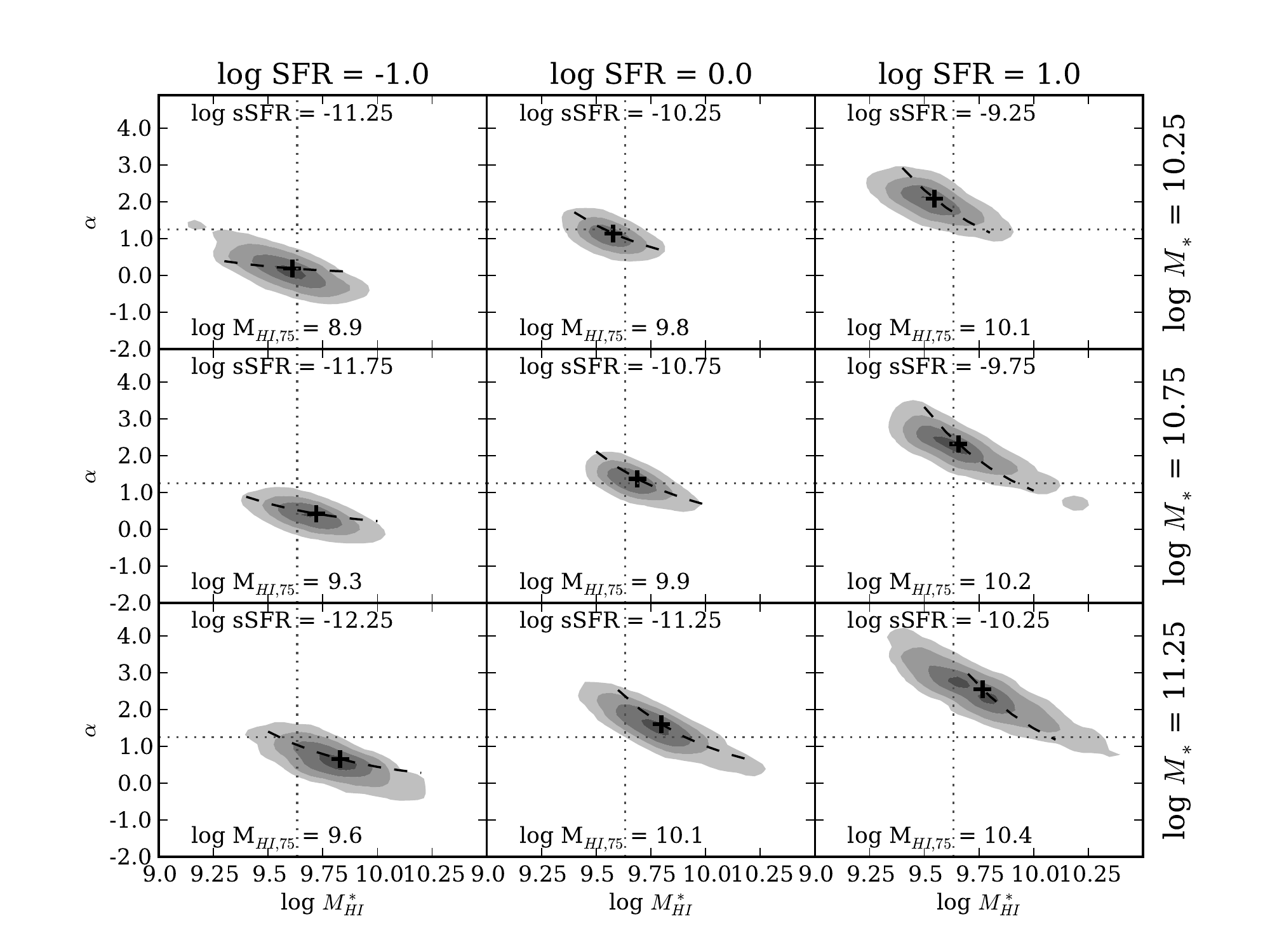}
\caption{Each panel shows the distribution of $\alpha$ vs. M$_{HI}^*$ calculated for a given stellar mass (increases down each column) and SFR (increases across each row) from the fit in Section \ref{sec:otherfits}.  The black cross indicates the median values in each panel and the black dotted lines show the median values for log SFR = 0 and log M$_*$/M$_{\odot}$ = 10.5.  Dashed lines are lines of constant M$_{HI,75}$.} 
\label{fig:sfrgrid1} 
\end{figure*}

\begin{figure*}[t]
\epsscale{1.0}
\plotone{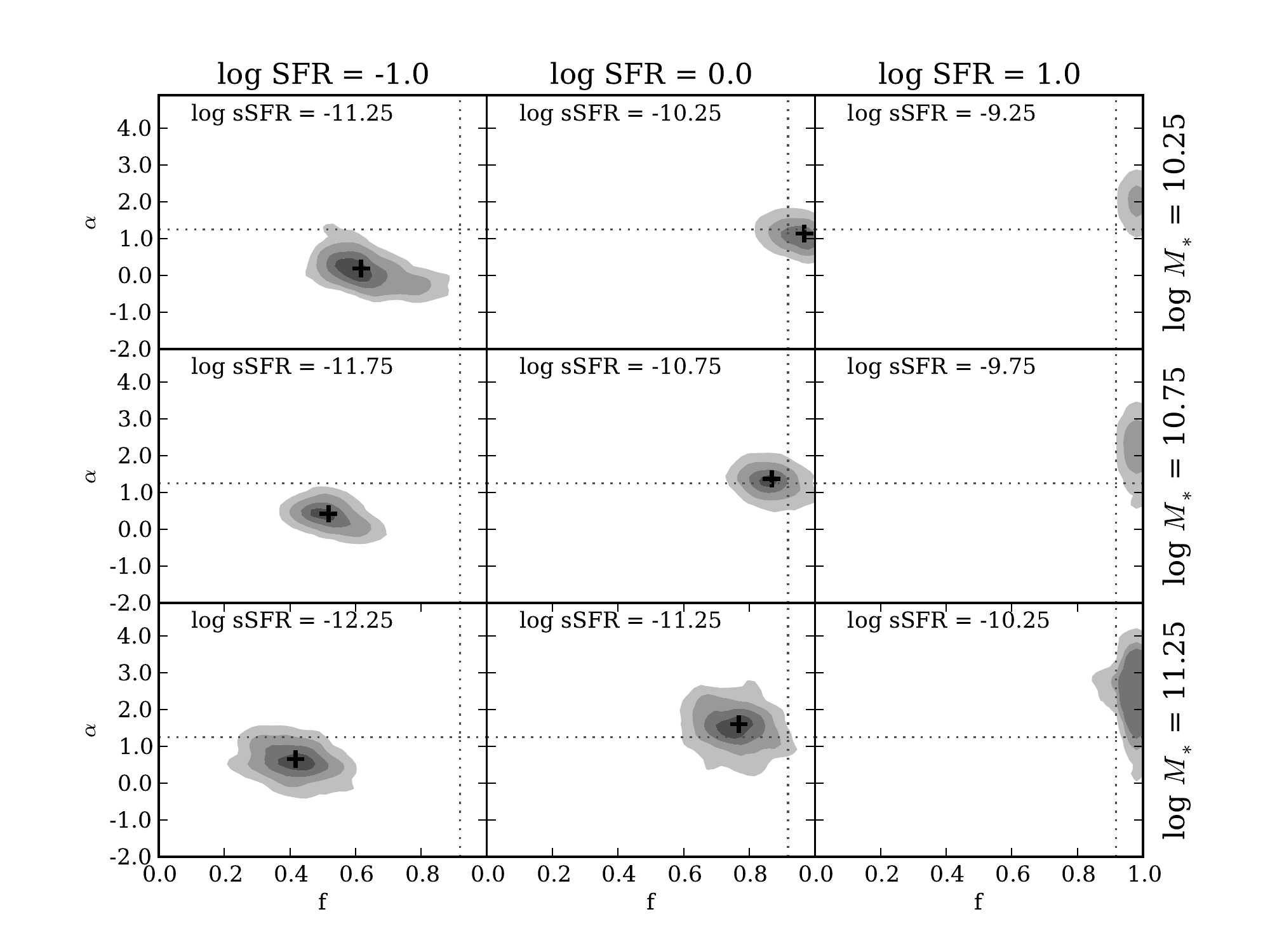}
\caption{Same as Fig. \ref{fig:sfrgrid1} for $\alpha$ and $f$.} 
\label{fig:sfrgrid2} 
\end{figure*}

In Figs. \ref{fig:sfrgrid1} and \ref{fig:sfrgrid2} we show the relationship between the three Schechter parameters and how they vary with 9 combinations of stellar mass and SFR that probe the GASS stellar mass range and 3 dex in SFR.  The contours are derived by using all 5000 iterations of the 9 MCMC parameters ($\alpha_o$, $\alpha_m$, $\alpha_s$, $M_o$, $M_m$, $M_s$, $f_o$, $f_m$, $f_s$) to calculate the 5000 sets of associated Schechter parameters ($\alpha$, M$_{HI}^*$, $f$) for each stellar mass-SFR bin.  The contours enclose 10, 25, 50, and 90\% of the 5000 calculated Schechter parameters.  SFR increases across the rows and stellar mass increases down each column.  

Fig. \ref{fig:sfrgrid1} reveals how the relationship between $\alpha$ and M$_{HI}^*$ changes with stellar mass and SFR.  As mentioned above, $\alpha$ increases with both stellar mass and SFR from a low of about 0.0 to a high of 2.5.  M$_{HI}^{*}$, on the other hand, shows less variation between 9.6 and 9.8.  This figure emphasizes the covariance between $\alpha$ and M$_{HI}^{*}$, also noted in Fig. \ref{fig:contours}.  

M$_{HI}^*$ denotes the transition point between the power law and the exponential regions of the Schechter function.  We might ask how a fixed quantile of the distribution varies and how this relates to M$_{HI}^*$.  We explored the analytic form of the relationship between M$_{HI}^*$ and M$_{HI,75}$ for a broken Schechter distribution and we found that the difference between the two quantities is a function of $\alpha$.  The curves in each panel of Fig. \ref{fig:sfrgrid1} show lines of constant M$_{HI,75}$, where the value of M$_{HI,75}$ is calculated assuming the median $\alpha$ and M$_{HI}^*$ in each bin (denoted by the crosshairs).  These lines follow the direction of covariance between $\alpha$ and M$_{HI}^*$.  Their covariance signifies the existence of a range of combinations of the two parameters that maintain a fixed M$_{HI,75}$ quantile.

While M$_{HI}^*$ varies by less than 0.3 dex across the range of stellar masses and SFRs represented here, M$_{HI,75}$ increases with stellar mass by at least 0.3 dex, and more so at low SFRs.  The variation in M$_{HI,75}$ shows that the quantiles of the distribution vary even though the peak at the characteristic HI mass remains relatively constant. 

Fig. \ref{fig:sfrgrid2} shows the same analysis of $\alpha$ and $f$.  $f$ ranges from 0.4 to 1.0 with this combination of stellar masses and SFRs and generally decreases with stellar mass and increases with SFR.  The negative correlation between $f$ and stellar mass was noted above.  
Values for $f$ cluster near 1.0 for galaxies with log SFR=1.0 because galaxies with moderate-to-high SFRs are very likely to have gas fractions above 1\%.  We constrained $f$ to not exeed 1.0 because that would be unphysical.

We can also examine how the Schechter parameters vary with specific SFR, noted in each panel of Figs. \ref{fig:sfrgrid1} and \ref{fig:sfrgrid2}.  As specific SFR decreases from the upper right corner to the lower left corner, the main difference is a strong decrease in $f$ - stronger than the variation in $f$ with respect to stellar mass or SFR alone.  Since $f$ is related to gas fraction, this means that specific SFR strongly depends on gas fraction.  Again, this result agrees with previous analyses \citep[e.g.][]{Catinella2010, Schiminovich2010}. 

\begin{figure*}[t]
\epsscale{1.0}
\plotone{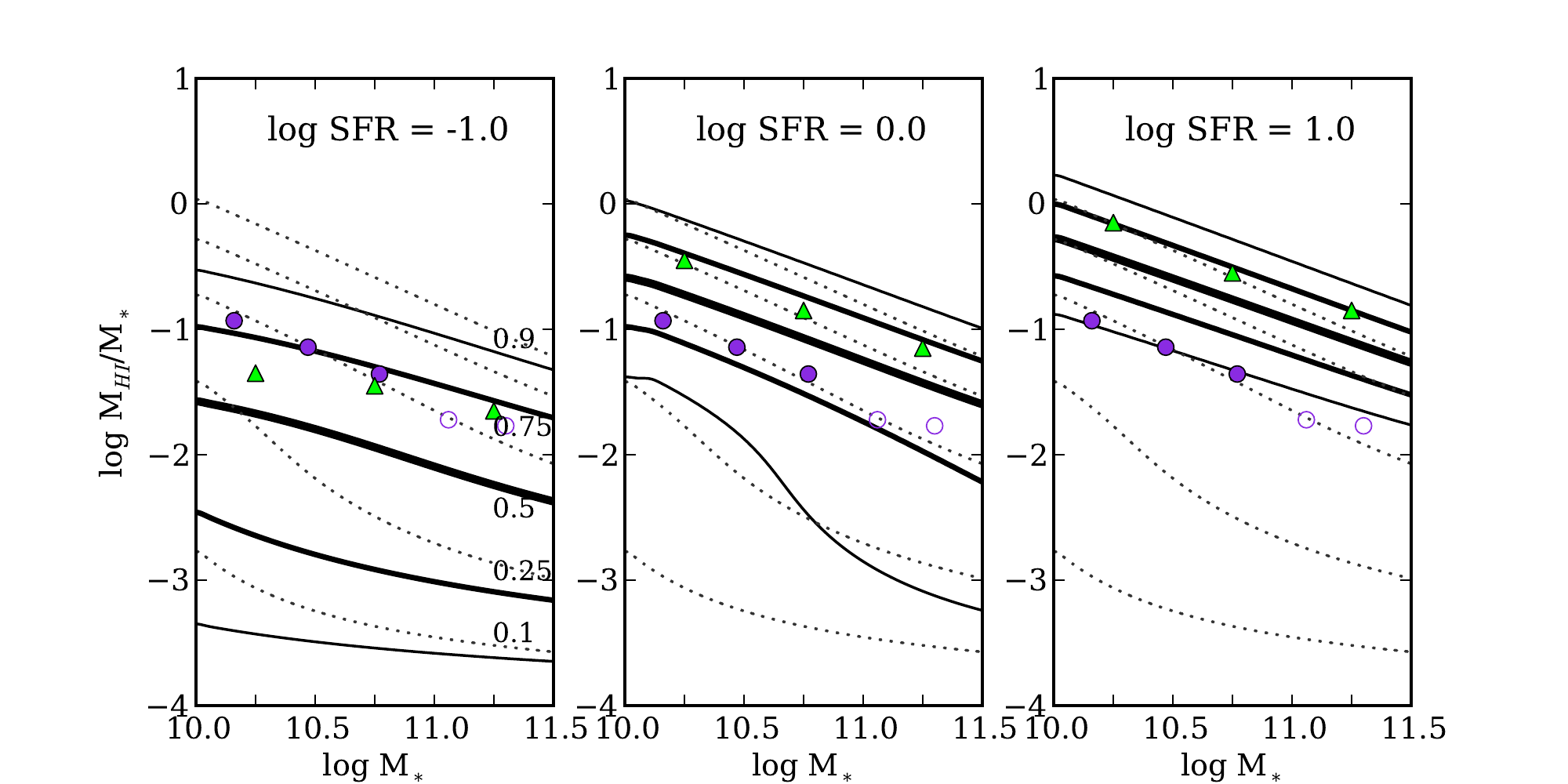}
\caption{Predicted HI gas fraction vs. stellar mass in three bins of SFR based on the continuous trivariate fit.  Solid curves indicate the fraction of galaxies with HI gas fractions below the lines.  The dotted curves show the same for the continuous bivariate fit.  Green triangles represent M$_{HI,75}$ in each M$_*$-SFR bin.  Blue circles show the median gas fraction (including upper limits) from \citet{Catinella2012}.  Empty blue circles are close to the gas fraction detection limit of GASS.}
\label{fig:triptych} 
\end{figure*}

Finally in Fig. \ref{fig:triptych} we show the distribution of HI gas fractions for three bins of SFR based on the higher dimensional fits examined above.  We used the bivariate and trivariate fits to calculate the distribution of HI masses in narrow bins of stellar mass and then divided by stellar mass to obtain the distribution of HI gas fraction.  Solid lines show quantiles based on the trivariate fit while dotted lines show quantiles based on the bivariate fit (the latter quantiles are the same for each SFR bin).  The trivariate fits show that the range of gas fractions varies significantly with SFR.  At log SFR = -1.0, galaxies with gas fractions in the middle 80\% of the distribution have a gas fraction range that spans 3 dex; at log SFR = 1.0, galaxies with gas fractions in the middle 80\% have gas fractions spanning only 1 dex.  Though the distribution of gas fractions changes with SFR, it doesn't change in proportion to SFR.  As SFR increases by a factor of 100, the gas content of galaxies with gas fractions in the top 10\% at a given SFR increases by less than a factor of 10.  The gas fraction vs. stellar mass distribution based on the bivariate fit exhibits a wide range of gas fractions and is most similar to the range of gas fractions for low-SFR galaxies in the trivariate fit.  Because the bivariate fit does not depend on SFR, it cannot capture the change in the gas fraction distribution with respect to SFR.  

We compare the gas fraction distribution based on these fits to two other measures of the gas fraction vs. stellar mass scaling relation.  First, we show that M$_{HI,75}$ calculated in each M$_*$-SFR bin tracks the trivariate fit's 75\% quantile fairly well in each SFR bin.  This fit matches M$_{HI,75}$ better than does the bivariate fit.  The better match to the trivariate fit confirms that SFR provides crucial additional information about HI content that is not already embedded in the relationship between HI mass and stellar mass.
 
To assess how our predicted gas fractions compare to median gas fractions calculated directly from the data, we show the median gas fraction scaling relation from \citet{Catinella2012}, which is averaged over galaxies of all SFRs.  The median gas fractions should agree well with the 50\% quantile, and they do if we look at the bivariate fit.  Turning to the trivariate fit, the \citet{Catinella2012} median gas fractions overestimate the gas fractions for galaxies with low SFRs and underestimate the gas fractions for galaxies with higher SFRs.

\begin{figure*}[t]
\epsscale{1.0}
\plotone{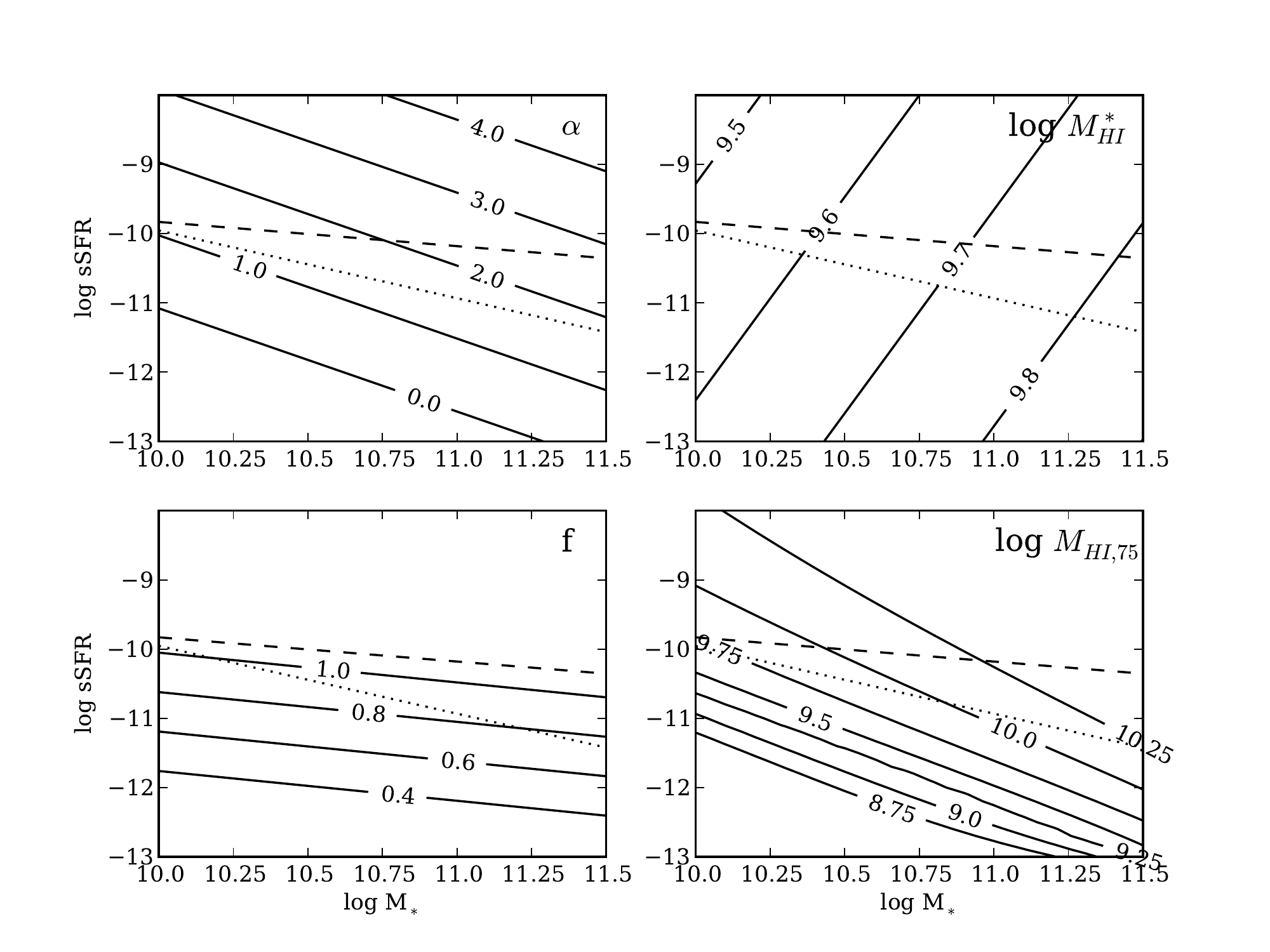}
\caption{Lines of constant $\alpha$, M$_{HI}^*$, $f$, and M$_{HI,75}$, calculated from the trivariate fit, in the specific SFR-stellar mass plane.  Dashed lines show the star-forming sequence from \citet{Salim2007}.  Dotted lines show the average specific SFR vs. stellar mass trend for GASS galaxies \citep{Schiminovich2010}.}
\label{fig:ssfr} 
\end{figure*}

To summarize, in Fig. \ref{fig:ssfr} we explore how the Schechter parameters vary within the specific SFR-stellar mass plane.  We find that $\alpha$ is high for galaxies with high stellar masses and high specific SFRs, which we also saw in Fig. \ref{fig:sfrgrid1}.  Lines of constant $\alpha$ are steeper than the star-forming sequence \citep{Salim2007} and average specific SFR vs. stellar mass trend for GASS galaxies \citep{Schiminovich2010}, indicating that $\alpha$ does have a dependence on stellar mass in addition to the dependence on SFR one would expect.  The upper right panel shows that M$_{HI}^*$ slowly increases with stellar mass and depends less on SFR.  Lines of constant $f$ confirm that galaxy populations with high specific SFR are likely to contain mostly galaxies with high gas fractions.  The lines of constant $f$ are almost parallel to the star-forming sequence, emphasizing the tight link between HI content and specific SFR.  Finally, we show lines of constant M$_{HI,75}$ in the last panel.  As we showed before, M$_{HI,75}$ exhibits a wider range of values than does M$_{HI}^*$.  Additionally, the lines for the two parameters are oriented almost perpendicular to each other.  While the peak of the distribution, M$_{HI}^*$, slowly increases with stellar mass, the shape of the distribution, indicated by the changing values of M$_{HI,75}$, depends strongly on both stellar mass and SFR.  Finally, the spacing between the lines of constant M$_{HI,75}$ change with SFR such that the value for M$_{HI,75}$ changes more quickly among highly star-forming galaxy populations.

To test the robustness of these trends, we used the 5000 iterations of the trivariate MCMC run to derive the standard deviation of $\alpha$, M$_{HI}^*$, $f$, and M$_{HI,75}$ at the center of each panel.  The standard deviations (0.26, 0.08, 0.04, and 0.03, respectively), which are smaller than the uncertainties on the individual parameters in the binned models and the continuous models (e.g. $\alpha_m$, $\alpha_s$, $\alpha_o$), are also small compared to the contour spacing so these trends are robust.

\subsection{Physical Drivers of the HIMF}

The fits presented above and in Section \ref{sec:himf} reveal how the HISMF varies quantitatively with stellar mass and SFR.  We can use these trends to better understand how the HISMF is shaped by various physical processes that act within different ranges of stellar mass and HI mass.  To guide our discussion of the physical drivers of the HISMF, we consider four aspects of the HISMF: 1) the steep dropoff at high HI masses; 2) the invariance of M$_{HI}^{*}$ with respect to stellar mass; 3) the dependence of $\alpha$ on stellar mass and SFR; and 4) the shape of the HISMF at low HI masses. 

\subsubsection{Steep Slope at High Masses}  
Our results show that it is very uncommon for galaxies to exist containing $>$ 10$^{10.5}$ M$_{\odot}$ of HI.  There exist several scenarios in which the buildup of HI may be stalled or halted.  Since M$_{HI}^{*}$ is largely independent of stellar mass and SFR, we can conjecture that the processes responsible for suppressing the buildup of HI are the same for most galaxies in the GASS stellar mass range.  To determine how various processes affect the HI-rich end of the HISMF, we turn to recent simulations that have incorporated interstellar medium physics in an attempt to reproduce the observed HIMF and understand its origin (though the difficulty of doing so is emphasized in \citet{Fontanot2013}).

\citet{Duffy2012}, \citet{Dave2013} and \citet{Kim2013} derived HIMFs from cosmological simulations with varying types and levels of feedback and compared them to the observed HIMF.  \citet{Duffy2012} found that their HIMF depended more strongly on their self-shielding prescriptions than a range of feedback prescriptions.  \citet{Kim2013} and \citet{Dave2013} (who also included self-shielding in their simulations) found that the variations in their feedback models affected the stellar mass or luminosity functions more than the HIMF.  In the \citet{Kim2013} models, strong supernovae feedback decreased the amplitude of the HIMF and luminosity function and steepened the slope of the HIMF at the HI-rich end.  AGN feedback, often used to suppress the growth of massive galaxies, also steepened the slope at the HI-rich end.  The outflow models in \citet{Dave2013} fit the HIMF better when their wind speed and mass-loading factor depended on galaxy velocity dispersion.  But their models require a quenching prescription to match the massive end of the stellar mass function; a more physical model could change the properties of massive and HI-rich galaxies, so the present results are difficult to interpret.

Is it possible that the steep slope at high HI masses results from high HI masses triggering the formation of H$_2$ or star formation and depleting HI?  For this scenario to be true, recently acquired gas must be funneled to the center of the galaxy where it is most likely to contribute to high HI surface densities that can fuel star formation.  The universal neutral hydrogen profile uncovered by \citet{Bigiel2012} indicates that this might happen in non-interacting spiral galaxies.  They found that the radial distribution of neutral gas (HI + H$_2$) surface density is remarkably similar across a sample of 33 nearby spirals.  They suggest that continuous gas inflow is responsible for this result by providing a fresh supply of neutral gas to the centers of galaxies where star formation depletes the neutral gas content.  Though recent simulations show that cold gas accreted via cosmological filaments tend to have high angular momentum and is deposited at large radii where the HI surface density is low \citep{Kimm2011, Stewart2011}, the resulting extended HI disks are likely short-lived as inflow will even out the distribution of cold gas.  
\subsubsection{The Peak at M$_{HI}^*$ and the Invariance of M$_{HI}^*$}
\label{sec:constantmstar}

The HISMF for massive galaxies has a feature that is not seen in the HIMF derived from surveys such as ALFALFA \citep{Martin2010}: the HISMF is peaked at M$_{HI}^*$ whereas the HIMF merely bends at M$_{HI}^*$ and has more galaxies with lower HI masses.  Why does the HISMF exhibit a peaked distribution?

\begin{figure*}[t]
\epsscale{1.0}
\plotone{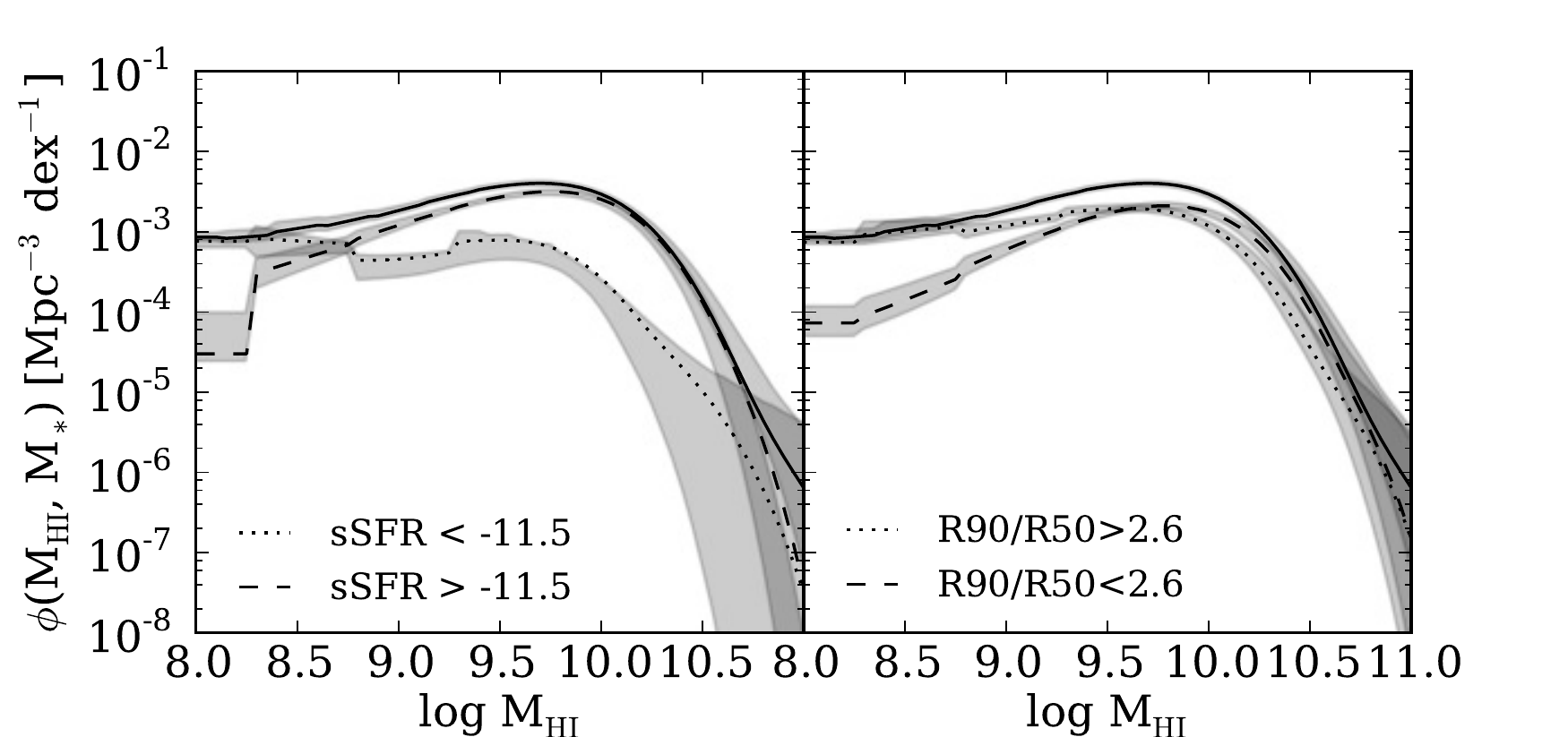}
\caption{The total HISMF divided by specific SFR and concentration (dashed and dotted lines) compared to the total HISMF (solid line).  \emph{Left:} The total HISMF for passively evolving (sSFR $<$ -11.5) galaxies (dotted) and star-forming (sSFR $>$ -11.5) galaxies (dashed).  \emph{Right:} The total HISMF for bulge-dominated galaxies (R$_{90}$/R$_{50}$ $>$ 2.6; dotted line); the total HISMF for disk-cominated galaxies (R$_{90}$/R$_{50}$ $<$ 2.6; dashed line).} 
\label{fig:colorcut} 
\end{figure*}

To evaluate whether the peak at M$_{HI}^*$ is dominated by star-forming galaxies with high HI masses, we derive the fit to the HISMF for subsamples of the GASS sample defined by SFR and concentration index.  In Fig. \ref{fig:colorcut} we show how galaxies with various SFRs and concentrations contribute to the total HISMF for massive galaxies.  As before, we fit a broken Schechter function to each SFR or concentration subset in several stellar mass bins and sum the normalized fits to obtain the total HISMF across the GASS stellar mass range.  To maintain a statistically significant number of galaxies in each stellar mass bin and in each subset, we define only two subsets with respect to SFR and concentration and we divide the sample into three evenly spaced stellar mass bins instead of six.  

In the left panel we divide the sample into two categories based on specific SFR: sSFR $<$ -11.5 and sSFR $>$ -11.5.  
Fig. \ref{fig:colorcut} shows that non-star-forming galaxies with sSFR $<$ -11.5 represent a large fraction of the massive galaxies with low HI masses (log M$_{HI}$/M$_{\odot}$ $\sim$ 8.0) and their contribution to the total HISMF decreases towards higher HI masses.  This is not surprising if one assumes that the reason for low SFRs is a lack of cold gas.  Star-forming galaxies, on the other hand, contribute less to the HISMF at low HI masses and represent many of the HI-rich galaxies with log M$_{HI}$/M$_{\odot}$ $\sim$ 10.0.  The peak of the total HISMF for massive galaxies is largely created by star-forming galaxies.

In the right panel we show the contribution to the HISMF from bulge-dominated (R$_{90}$/R$_{50}$ $>$ 2.6) and disk-dominated (R$_{90}$/R$_{50}$ $<$ 2.6) galaxies.  (The dividing line is taken from \citet{Strateva2001}.)  Though bulge-dominated galaxies tend to have lower HI masses and disk-dominated galaxies tend to have higher HI masses, the difference between bulge-dominated and disk-dominated galaxies is less pronounced than the difference between star-forming and non-star-forming galaxies, especially at high HI masses.  This points to specific SFR as a better indicator of cold gas content than the presence of a bulge.  Indeed, GASS has shown that NUV-r color, a proxy for specific SFR, is a much better predictor of HI gas fraction than concentration index \citep{Catinella2010, Fabello2011a, Catinella2012}.

Together, these two cuts can inform our interpretation of the shape of the HISMF and the invariance of M$_{HI}^*$.  Because star-forming and disk-dominated galaxies dominate the HI-rich end of the HISMF, they appear to drive the shape of the peak of the distribution.  Passively evolving, centrally concentrated galaxies exhibit no such peak at M$_{HI}^*$ and tend to lie at the HI-poor end of the HISMF.  They will have a more significant impact on $\alpha$, the slope of this part of the HISMF.  We will discuss this in more detail below.

We can also refer to simulations to see if they reproduce a peaked HIMF.  \citet{Lagos2011b} show that low \emph{halo} mass bins exhibit peaked HIMFs, though it is not clear if the existence of a peak in our HISMF is dependent on galaxy mass in the same way.  \citet{Kauffmann2012} recreate peaked HI gas fraction distribution functions in bins of stellar mass, stellar mass surface density and concentration with their semi-analytic models, but their models fail to reproduce other important observational trends.

We find that M$_{HI}^*$, the HI mass at the peak of the HISMF, is nearly constant with respect to stellar mass despite its dependence on SFR.  (The near constant HI mass for galaxies in the GASS mass range was noted previously \citep{Catinella2010}.)  Although SFR is known to increase with stellar mass within the population of star-forming galaxies \citep[e.g.][]{Salim2007}, the average SFR for the GASS sample, which is representative of massive galaxies in the local universe, is constant with respect to stellar mass (see dotted lines in Fig. \ref{fig:ssfr}).  M$_{HI}^*$ can be constant with respect to stellar mass because the average SFR and the star formation efficiency (SFE = SFR/M$_{HI}$) are nearly constant within this mass range \citep{Schiminovich2010}.

A possible explanation for the near constant M$_{HI}^*$ is that galaxies tend to maintain an equilibrium HI mass even as they form stars and their stellar masses increase.  If one considers a galaxy's HI mass as a product of the competing processes of gas accretion and gas consumption, a constant HI mass suggests that these processes cancel each other out within individual galaxies, maintaining a similar HI mass distribution across a range of stellar masses.  Indeed, simulations have shown that for star-forming galaxies, the mass inflow rate is generally balanced by the sum of the mass outflow rate and the SFR, which also removes gas from the ISM \citep[Eq. 1 in][]{Dave2012, Schaye2010, Lagos2011a}.  \citet{Hopkins2012} showed that star formation is regulated by stellar-driven winds whose mass-loading increases with SFR.  \citet{Dave2012} explained that galaxies out of equilibrium tend to return to equilibrium, such as when a galaxy that receives an influx of gas in a large accretion event subsequently experiences a higher SFR triggered by the higher gas fraction.  Thus the increased gas consumption in the form of star formation balances the increased accretion rate and the galaxy returns to equilibrium.  An alternative explanation for the constant M$_{HI}^*$ - that HI masses of galaxies vary in unpredictable ways as they evolve but that the distribution of HI masses is somehow maintained - is unlikely.  Thus, our results support a scenario in which the processes that contribute to and deplete the HI content of a galaxy conspire to maintain an equilibrium HI mass. 

Finally, we note that \citet{Prochaska2009} also uncovered a universal distribution of HI content: they found that damped Ly$\alpha$ systems exhibit the same HI column density distributions between z=2.2 and z=5, and that the distributions at these high redshifts matches the distribution function for HI disks in the local universe.  Their explanation that various processes affecting the HI content of galaxies must affect the inner, high-column density regions and the outer, low-column density regions of galaxies similarly could also apply to our result.

\subsubsection{The Dependence of $\alpha$ on Stellar Mass and SFR}

The parameter $\alpha$ measures the slope of the HISMF between M$_{HI,break}$, which is a function of stellar mass, and M$_{HI}^*$, which depends weakly on stellar mass.  Thus, the range of HI masses that $\alpha$ describes becomes much narrower at higher stellar masses and the number of galaxies contributing to the estimate of $\alpha$ decreases.  Our definition of $\alpha$ contrasts with standard treatments of $\alpha$, in which there is no cutoff at low masses and $\alpha$ describes the low-mass end of mass functions.  We do not compare our derived values of $\alpha$ to other published values since we use $\alpha$ to characterize the shape of the HIMF at more moderate HI masses.  We also note that in our model $\alpha$ represents the entire exponent in the Schechter function, whereas other authors have used $\alpha+1$ as the exponent \citep[e.g.][]{Martin2010}.

We find that $\alpha$'s dependence on stellar mass and SFR is stronger than the other parameters' dependencies.  The wide range of values for $\alpha$ is illustrated in Fig. \ref{fig:sfrbins} and significantly impacts the shape of the HISMF.  If $\alpha$ is close to zero, then the HISMF has a relatively flat distribution below M$_{HI}^*$.  For higher values of $\alpha$, the distribution above M$_{HI,break}$ becomes strongly peaked about M$_{HI}^*$.  

$\alpha$ increases with both stellar mass and SFR causing samples with massive galaxies and with highly star-forming galaxies to have peaked distributions of HI masses.  Assuming that high SFRs are in part driven by the presence of large amounts of cold gas, it makes sense that there would exist fewer galaxies with low HI masses within a population of star-forming galaxies.  The HISMF is flat or has a slowly decreasing slope at moderate HI masses when galaxies with low stellar masses or low SFRs comprise the population.  The different distributions of HI masses at various SFRs has implications for the relationship between HI and star formation.  Galaxies with low SFRs can have a wide range of HI masses while highly star-forming galaxies are more likely to have a high HI mass.  This was also evident in Fig. \ref{fig:triptych}.  Thus, galaxies with low SFRs do not have a well-defined relationship between HI and star formation while galaxies with high SFRs exhibit a much tighter relationship between HI and star formation.  

The positive correlation between $\alpha$ and stellar mass for the bivariate and trivariate fits shows that HI masses are more strongly peaked among samples of massive galaxies independent of their SFRs.  This trend shows that at high stellar masses galaxies tend to have higher HI masses as well.

Although there does not exist an observational HISMF for comparison, we can look to \citet{Springob2005}, who found that the HIMF depends on morphology.  In particular, they noted that the HI-poor end of the HIMF is close to flat (or $\alpha$ is close to zero using our convention) when only early-type spirals are considered.  The HIMF for later-type spirals rises towards lower HI masses ($\alpha$ is smaller and negative).  We find that the HISMF rises towards lower HI masses for galaxies with lower stellar masses.  Since later-type galaxies have, on average, lower stellar masses than early-type galaxies, it is likely that the same relationship between HI mass, stellar mass, and morphology is driving both of these trends.

\subsubsection{The Distribution of the HIMF at Low HI Masses}
\label{sec:lowHImass}

There are two aspects of the HI-poor end of the HISMF that must be examined: the number of galaxies (or fraction of galaxies with a given stellar mass) that have a low gas fraction and the distribution of these galaxies with respect to HI mass.  We quantified the former with $1-f$ and found that the fraction of galaxies with HI gas fractions less than 1\% increases with stellar mass.  If we assume that all galaxies are undergoing infall proportional to their halo masses \citep[e.g.][]{Dekel2013} then we must identify the processes that remove gas from galaxies to understand what sets $1-f$ and why it changes with stellar mass.  These processes likely include AGN feedback, which heats gas in the disk and surrounding halo \citep[e.g.][]{Croton2006, Hopkins2008, Gabor2011}, environmental effects that deplete gas \citep[e.g.][]{Tonnesen2007}, and stellar-driven winds that expel gas from galaxies \citep[e.g.][]{Oppenheimer2008, Oppenheimer2010}.  The GASS sample does not include many galaxies in dense environments so we ignore the effect of environment here, though we note that the stripping of gas can have a significant effect on galaxies in clusters.  Likewise, stellar-driven winds do not remove large amounts of gas from massive galaxies because the gas can be reaccreted in the form of a galactic fountain \citep{Oppenheimer2008, Oppenheimer2010}.  It is likely that these processes as well as star formation contribute to the HI-poor end of the HISMF by removing some cold gas from galaxies or preventing gas from cooling.  Though simulations have shown that such processes can effectively remove cold gas from galaxies, they do not always predict the right number of HI-poor galaxies.  \citet{Dave2013} predict the distribution of HI gas fractions as a function of stellar mass and find a negligible number of massive galaxies with gas fractions below 1\%, in contrast to our HISMF, which predicts that 20-50\% of galaxies in this stellar mass range should have such low gas fractions.

The precise distribution of massive galaxies with low gas fractions is unknown and we do not attempt to fit the distribution of HI masses below M$_{HI,break}$, where the HI measurements are mostly upper limits.  Instead, we show the predicted space density of galaxies below a 1\% gas fraction as a constant distribution with respect to stellar mass.  Here we discuss whether that decision makes sense physically.  The distribution of galaxies at low HI masses depends, in part, on the ways in which galaxies with little or no HI reacquire small amounts of cold gas.  If we assume that a given fraction of halo baryons cool onto the disk from the hot halo \citep{Anderson2010, Anderson2013}, then we might expect the distribution of HI masses to exhibit a peak at low HI masses that depends on halo mass.  However, stochastic additions to the cold gas in the disk, such as from infalling satellites or stellar mass loss \citep{Leitner2011} would likely smooth the distribution.

There is some indication that galaxies are distributed uniformly at low HI masses.  \citet{Serra2012} measure the HI masses and characterize the HI morphology of 166 early-type galaxies brighter than M$_K$ = -21.5.  They find that galaxies with HI that they describe as ``unsettled" (i.e. HI in streams or tails rather than an ordered disk) can contain a wide range of HI masses from a few times 10$^7$ to 10$^{10}$ M$_{\odot}$ whereas galaxies with settled HI disks tend to have higher HI masses.  Galaxies with unsettled HI might be accreting small amounts of cold gas episodically, such as from the halo \citep{Putman2012} or intergalactic medium \citep{Keres2009b}, and might not have a preferred HI mass scale.

\citet{Serra2012} derive an HIMF for early-type galaxies, which is relatively flat for 7 $<$ log M$_{HI}$/M$_{\odot}$ $<$ 9.5, and an HIMF for spirals, which is sharply peaked and falls off sharply on either side of M$_{HI}^*$.  If late-type galaxies are the main contributors to the peak of the HIMF, as the \citet{Serra2012} results and our Fig. \ref{fig:colorcut} suggest, then it is possible that early-type galaxies, and others undergoing episodic gas acretion, contribute to a broad distribution of HI masses at the HI-poor end of the HIMF.

But there is some evidence that the HISMF for massive galaxies instead rises at low HI masses, making the HISMF for massive galaxies bimodal.  We already noted above that this trend is evident in the simulated HIMF for galaxies in bins of halo mass from \citet{Lagos2011b}.  It is also reasonable to conjecture that the HISMF might mimic the distribution of SFRs in bins of stellar mass.  Observations reveal that the distribution of SFRs has a star-forming sequence plus a tail towards passively evolving galaxies at a wide range of stellar masses \citep{Salim2007, Wyder2007, Schiminovich2007} and some simulations have begun to reproduce this distribution by varying star formation prescriptions \citep{Lagos2011a}.  Though it is difficult to make strong conclusions about the HI-poor end of the HIMF from surveys that are flux-limited, the \citet{Serra2012} HIMF for early-type galaxies reveals that the HI-poor end could slope upward if the HI in undetected galaxies is included.  Finally, the results from our trivariate fit show that at the lowest stellar masses and SFRs we probe, $\alpha$ could become negative, indicating a possible peak at low-to-intermediate HI masses (see bold line in Fig. \ref{fig:sfrbins}).  

\section{Summary and Conclusions}

We have used 480 galaxies from the GALEX Arecibo SDSS Survey \citep[GASS;][]{Catinella2010} Data Release 2 \citep{Catinella2012} that were observed in HI at Arecibo to derive the bivariate HI mass-stellar mass function for massive galaxies with log M$_{*}$/M$_{\odot}$ $>$ 10.  Below we summarize our key results:

\begin{itemize}  

\item We use an MCMC routine to fit six different models, three variations of the Schechter function and three variations of the log-normal function, to the HISMF in six bins of stellar mass from log M$_{*}$/M$_{\odot}$ = 10.0 to log M$_{*}$/M$_{\odot}$ = 11.5.

\item The Schechter and log-normal parameters M$_{HI}^*$, $\alpha$, $\mu$, and $\sigma$ show little variation with stellar mass, though $f$, the fraction of galaxies with gas fractions above 1\%, decreases from about 80\% to 40\%.  The continuous bivariate fit shows that $\alpha$ varies as M$_*^{0.39}$, M$_{HI}^*$ varies as M$_*^{0.07}$ and $f$ varies as M$_*^{-0.24}$.  The continuous bivariate fit should be taken as our main result to compare to future observations and simulations.  In particular, we believe that the change in $f$ with stellar mass will provide a strong constraint for simulations.

\item To test the accuracy of the models, we compare the total HISMF for massive galaxies, constructed by summing the HISMF across all stellar mass bins, to the data.  This comparison shows that each variation of the log-normal function overestimates the number of HI-rich galaxies.  The three Schechter functions match the data reasonably well, though we choose to proceed only with the broken Schechter function, which extends to a 1\% gas fraction with a fractional contribution below that limit.

\item The total HISMF for massive galaxies is consistent with the ALFALFA HIMF \citep{Martin2010} at high HI masses, indicating that the most HI-rich galaxies in the local universe are also massive in stars.  We find that massive galaxies contribute 41\% of the total HI density in the local universe.

\item To understand the physical drivers of the shape of the HISMF, we derive the continuous trivariate HI-M$_*$-SFR function, for which we redefine the Schechter parameters as functions of stellar mass and SFR.  Though the bivariate fit uncovers an HISMF that varies only weakly with stellar mass, the trivariate fit shows that the shape of the HIMF is a strong function of SFR.  The trivariate fit shows that $\alpha$ varies as M$_*^{0.47}$ and SFR$^{0.95}$.

\item We show that the peak at M$_{HI}^*$ is likely dominated by star-forming galaxies and that its slow variation with stellar mass could be related to the ways in which star-forming galaxies maintain an equilibrium HI mass.

\item The dependence of $\alpha$ on SFR shows that highly star-forming galaxies tend to have a narrow range of HI masses peaked about a high HI mass whereas passively evolving galaxies can have a wider range of HI masses.

\end{itemize}

The bivariate HI mass-stellar mass function provides a two-dimensional description of the ways in which HI mass and stellar mass are distributed among massive galaxies.  The trivariate HI-M$_*$-SFR function expands upon this by exploring its relationship to SFR.  These are important constraints for simulations; likewise, simulations can shed light on the shape of these distribution functions by testing how feedback and star formation prescriptions affect them.  For example, work done by \citet{Dave2013} suggests that galactic outflows could play an important role in preventing the buildup of HI in massive galaxies.  Future simulations should begin to explore these bivariate and trivariate functions now that they have begun to incorporate interstellar medium physics and have had success matching observed one-dimensional distribution functions.  Future observations can expand this work by extending these trends to lower stellar masses, a goal that is achievable with current facilities, and to higher redshifts, which should be possible with the next generation of radio telescopes.

\acknowledgments

The authors wish to thank David Hogg for useful discusssions and Dan Foreman-Mackey for suggestions that improved this paper.  The authors also thank the anonymous referee for valuable comments.

The Arecibo Observatory is operated by SRI International under a cooperative agreement with the National Science Foundation (AST-1100968), and in alliance with Ana G. M\`{e}ndez-Universidad Metropolitana, and the Universities Space Research Association.

\emph{GALEX (Galaxy Evolution Explorer)} is a NASA Small Explorer, launched
in April 2003. We gratefully acknowledge NASA's support for
construction, operation, and science analysis for the \emph{GALEX} mission,
developed in cooperation with the Centre National d'Etudes Spatiales
(CNES) of France and the Korean Ministry of Science and Technology.

This work has made extensive use of the \url[]{MPA/JHU}
SDSS value-added catalogs.

Funding for the SDSS and SDSS-II has been provided by the Alfred
P. Sloan Foundation, the Participating Institutions, the National
Science Foundation, the U.S. Department of Energy, the National
Aeronautics and Space Administration, the Japanese Monbukagakusho, the
Max Planck Society, and the Higher Education Funding Council for
England. The SDSS Web Site is http://www.sdss.org/.

The SDSS is managed by the Astrophysical Research Consortium for the
Participating Institutions. The Participating Institutions are the
American Museum of Natural History, Astrophysical Institute Potsdam,
University of Basel, University of Cambridge, Case Western Reserve
University, University of Chicago, Drexel University, Fermilab, the
Institute for Advanced Study, the Japan Participation Group, Johns
Hopkins University, the Joint Institute for Nuclear Astrophysics, the
Kavli Institute for Particle Astrophysics and Cosmology, the Korean
Scientist Group, the Chinese Academy of Sciences (LAMOST), Los Alamos
National Laboratory, the Max-Planck-Institute for Astronomy (MPIA),
the Max-Planck-Institute for Astrophysics (MPA), New Mexico State
University, Ohio State University, University of Pittsburgh,
University of Portsmouth, Princeton University, the United States
Naval Observatory, and the University of Washington.

\appendix
\section{Applications of the HIMF:  Comparing to Simulations and Photometric Gas Fractions}

The HISMF provides a crucial constraint for estimates of HI content for populations of observed or galaxies.  Simulations are generally normalized and adjusted until their output matches the observed one-dimensional stellar mass function and/or HIMF \citep{Lagos2011b, Duffy2012, Dave2013, Kim2013}.  The HISMF adds an additional constraint because it defines not only the stellar mass function or HIMF but the relation between the two.   Indeed, a sophisticated comparison between observations and simulations necessitates the use of the HISMF.  For example, \citet{Lagos2011b} demonstrate that galaxies in their simulations occupy the same region of the HI gas fraction - stellar mass plane as the GASS galaxies do by comparing their distribution of simulated z=0 galaxies to individual GASS detections and non-detections.  It would be more revealing to ascertain whether the simulations and the observations yield samples that not only lie in the correct range of parameter space but are distributed within that parameter space in a similar manner.  The HISMF derived here makes that comparison a possibility.

Though the most obvious application of the HISMF is to simulations, it can be applied to a range of problems including photometric gas fractions.  The difficulty of measuring HI in large unbiased samples of galaxies has led to the development of photometric gas fraction relations, which provide estimates of gas fractions are based on established correlations between HI gas fraction and other physical parameters such as color, surface brightness, and stellar mass surface density.  A comparison between the HIMFs derived from various photometric gas fraction estimates and the HISMF derived here is a useful test of their validity.

We compile photometric gas fraction relations from the literature and compare their HISMFs to our HISMF.  We select photometric gas fraction relations from \citet[][their Fig. 1b]{Kannappan2004}, \citet[][Eq. 4]{Zhang2009}, \citet[][their Fig. 12]{Catinella2010}, \citet[][Eq. 5]{Cortese2011}, and \citet[][Eqs. 2 and 5] {Huang2012}.  We apply them to all galaxies in the GASS parent sample and derive corresponding HI masses to be used in the HISMF fit.  Though photometric gas fraction relations can be applied to galaxies with a wide range of parameters, \citet{Cortese2011} suggested that they might be valid only for star-forming galaxies.  To account for this, we apply our standard treatment of detections and non-detections to the derived photometric gas masses where galaxies with NUV-r $<$ 4 are treated as detections and photometric gas masses for galaxies with NUV-r $>$ 4 are treated as upper limits of non-detections.
  
\begin{figure*}[t]
\epsscale{1.0}
\plotone{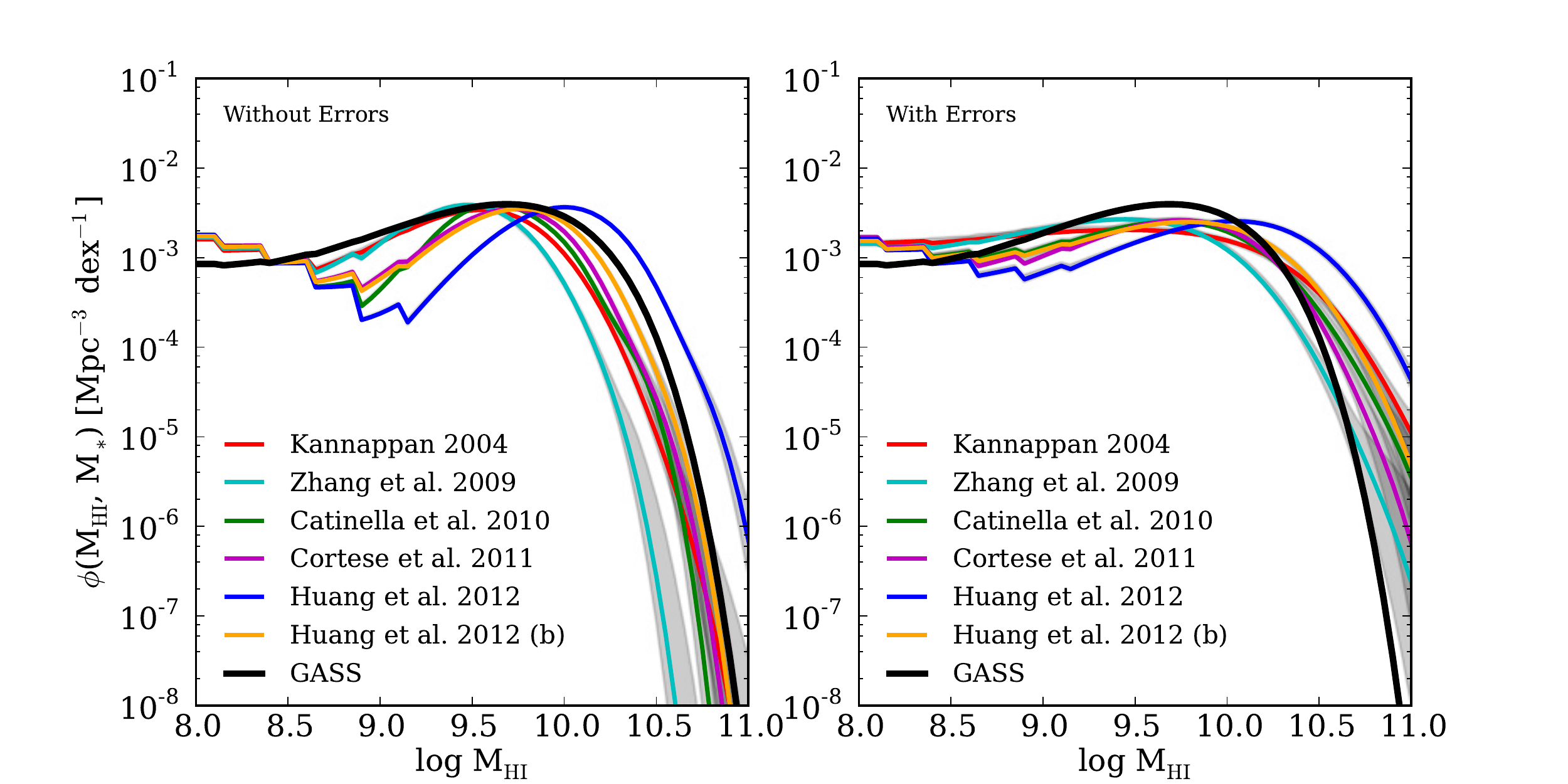}
\caption{Broken Schechter fits to the total HISMF derived from published photometric gas fractions calculated for galaxies in the GASS parent sample.  The fits presented in the left panel do not include errors while the fits shown in the right panel do include errors.  The black line shows the total GASS HISMF based on the continuous bivariate fit; the colored lines show the total HISMF for massive galaxies based on photometric gas fractions.  Because the best-fit models to the distributions of photometric gas fractions are poor in the highest stellar mass bin (see Figs. \ref{fig:phot_all_1} and \ref{fig:phot_all_2} below), here we show the sum of the HISMF only up to log M$_*$/M$_{\odot}$ = 11.25.}
\label{fig:phot} 
\end{figure*}
 
For simplicity, we first compare the total HISMFs derived from photometric gas fractions to our total continuous bivariate HISMF in Fig. \ref{fig:phot}.  We calculate the photometric HIMSFs with and without errors on the predicted gas fractions and show the results of both sets of fits.  To include the errors on the photometric gas fractions, we add a value to the predicted gas fraction that is randomly selected from a Gaussian distribution whose width is equal to the reported error in each photometric gas fraction relation.  (\citet{Huang2012} do not report the scatter in their relations so we use an error of 0.3 dex, consistent with the most recently published photometric gas fraction relations.)  \citet{Li2012} show that the errors on the predicted gas fractions can significantly affect results when compared to observed HI gas fractions.

In the left panel of Fig. \ref{fig:phot} we compare the photometric HISMFs without errors to our total HISMF.  The photometric gas fraction relations generally underestimate the number of HI-rich galaxies.  \citet{Huang2012} presented two photometric gas fraction relations, both of which we present here.  The first is based on color and stellar mass surface density (their Eq. 2.) and the second uses color and the stellar mass surface density of the disk only (their Eq. 5) because HI is likely linked to the galaxy disk.  \citet{Huang2012} show that photometric gas fractions derived from this equation exhibit less scatter when compared to measured HI masses.  The HISMF derived using the disk photometric gas fraction is the only photometric HISMF that slightly overpredicts the number of HI-rich galaxies.  Though this result suggests that taking into account the disk mass represents a possible way to improve photometric gas fractions, the HISMF derived from this photometric gas fraction relation underestimates the number of galaxies with moderate (log M$_{HI}$/M$_{\odot}$ $\sim$ 10) HI masses.  Other photometric gas fraction relations are a better match to the distribution of HI-poor galaxies.

When we include in the fits the scatter in the photometric gas fraction relations, the comparison between the photometric and observed HISMFs is quite different: most photometric gas fractions overestimate the number of HI-rich galaxies and all underestimate the number of galaxies with moderate HI masses.

\begin{figure*}[t]
\epsscale{1.2}
\plotone{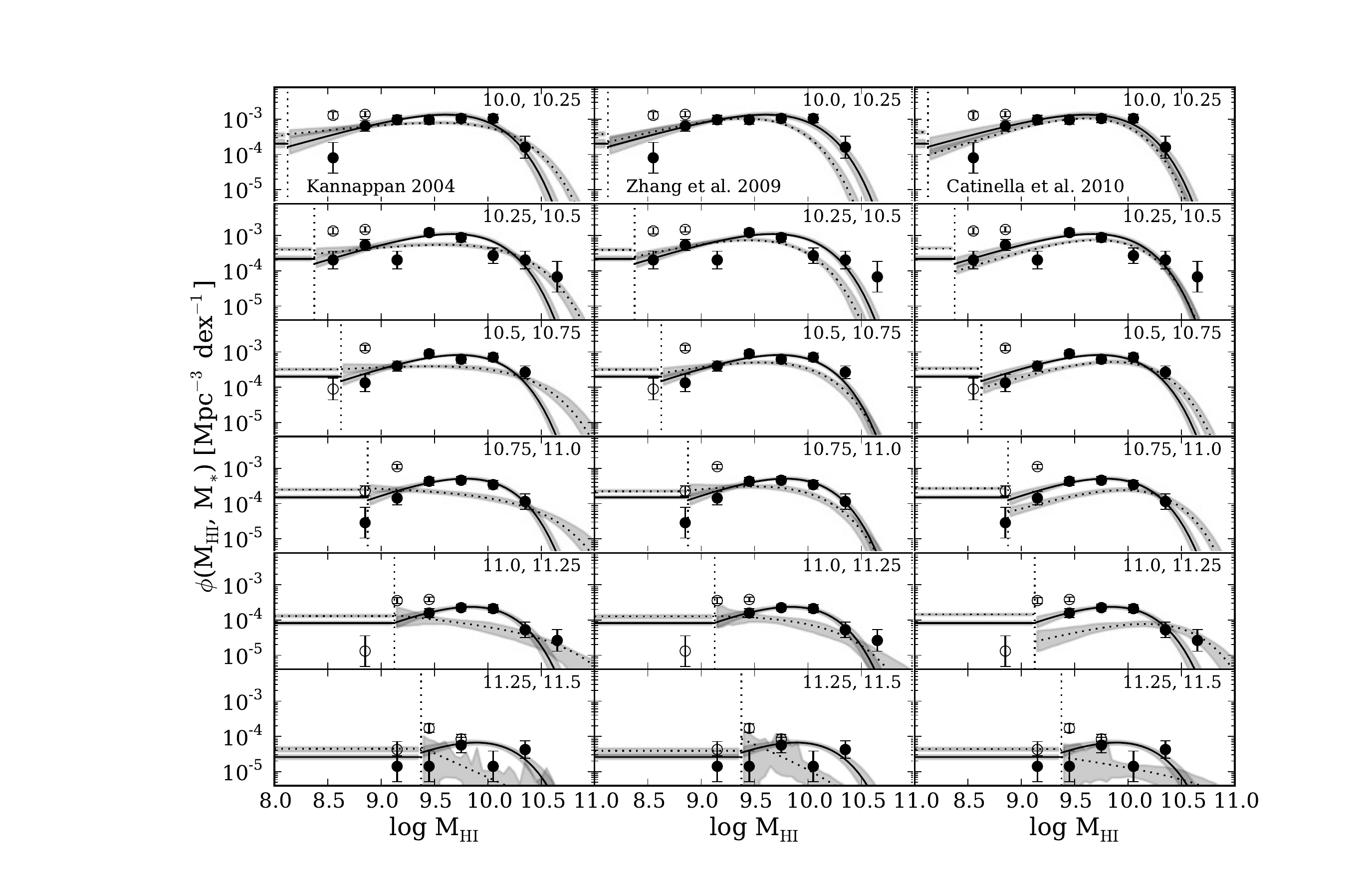}
\caption{The broken continuous bivariate Schechter fit to the GASS HISMF (solid lines) compared to the broken Schechter fit to the HISMFs derived from photometric gas fraction relations published in \citet{Kannappan2004}, \citet{Zhang2009}, and \citet{Catinella2010} (dotted lines).} 
\label{fig:phot_all_1} 
\end{figure*}

\begin{figure*}[t]
\epsscale{1.2}
\plotone{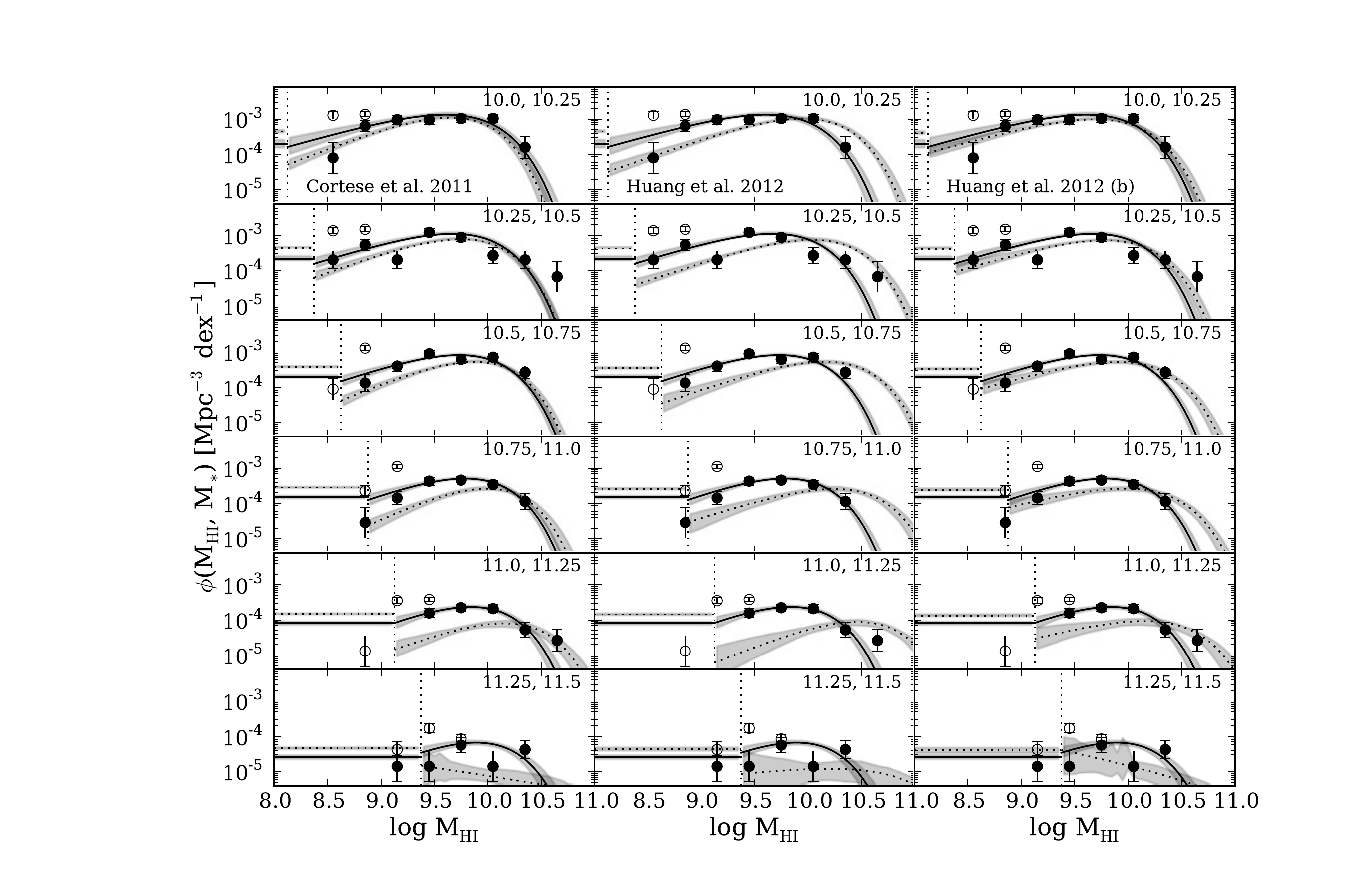}
\caption{Same for Fig. \ref{fig:phot_all_1} for the photometric gas fraction relations published in \citet{Cortese2011} and \citet{Huang2012}.} 
\label{fig:phot_all_2} 
\end{figure*}

There are several possible reasons why photometric gas fractions do not capture the details of the total HISMF for massive galaxies.  One explanation is that the small number of HI detections for HI-poor galaxies and the most massive galaxies hinders the ability to derive accurate photometric gas fraction relations across a wide range of HI masses.  Another explanation is that photometric gas fraction relations do not apply to non star-forming galaxies.  \citet{Cortese2011} argue that a plane describing the relationship between gas fraction, color, and stellar mass surface density may not be appropriate for use with these galaxies because they do not exhibit the same linear trends between HI gas fraction and both color and stellar mass surface density as star-forming galaxies.  This was also seen in \citet{Catinella2010}.  \cite{Zhang2009} found that their photometric gas fractions reproduce the observed HIMF from \citet{Zwaan2005} if they only consider star-forming galaxies, but we showed in the previous section that bulge-dominated galaxies and red galaxies do contribute significantly to the HISMF for massive galaxies at the HI-rich end.  

The \emph{binned} photometric and observed HISMFs provide a more nuanced view of the discrepancy between the photometric and observed HISMFs.  In Figs. \ref{fig:phot_all_1} and \ref{fig:phot_all_2} we present the binned HISMF derived for each of the six photometric gas fraction relations with errors.  Instead of analyzing the differences between each mass function in each stellar mass bin, we take a global view and note that the photometric HISMFs tend to diverge from the GASS HISMFs at higher stellar masses.  That the accuracy of photometric gas fractions varies with stellar mass is not surprising given the results of the trivariate fit in Section \ref{sec:otherfits}.  We showed that the shape of the HISMF changes with stellar mass and SFR.  Thus, a single photometric gas fraction estimator that assumes a simple relationship between gas content and other physical quantities is not likely to be valid across a wide range of stellar masses.  There is still work to be done in deriving photometric gas fraction relations that reproduce the true distribution of HI masses while also yielding accurate estimates of gas fractions for individual galaxies.  It is possible that an approach that considers how estimates of gas fraction might change with stellar mass will yield more accurate results.

\section{Assumed stellar mass function}

\begin{figure}[t]
\epsscale{0.5}
\plotone{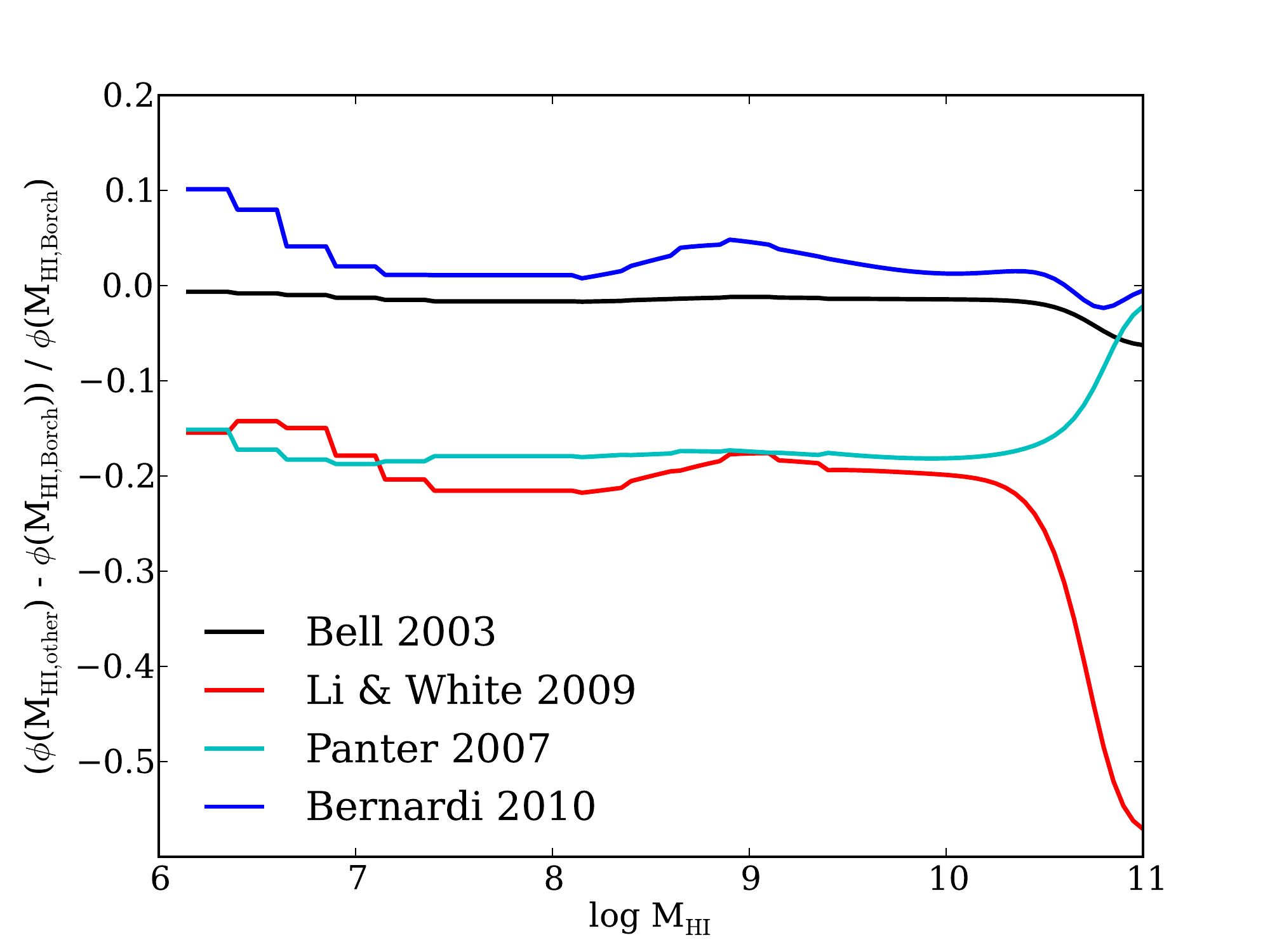}
\caption{The fractional difference in the total HISMF when stellar mass functions other than that in \citet{Borch2006} are used.} 
\label{fig:stellarmassfunctions} 
\end{figure}

We normalized the GASS HISMF by assuming the space density of galaxies in each stellar mass bin is given by the stellar mass function in \citet{Borch2006}.  Here we show that using other stellar mass functions would have a negligible effect on our final results.

We compiled z=0 stellar mass functions from \citet{Bell2003_SMF}, \citet{Panter2007}, \citet{Li2009} and \citet{Bernardi2010}.  (A comparison of these mass functions can be found in \citet{Bernardi2010}.)  They represent a range of approaches and techniques.  For example, \citet{Li2009} found that a triple Schechter function is a better match to the data than a single Schechter function.  \citet{Bernardi2010} included the effect of measurement errors to obtain a fit that matches the observed distribution of stellar masses rather than the intrinsic distribution of stellar masses.  

In Fig. \ref{fig:stellarmassfunctions} we show how our derived total HISMF would change if we assumed different stellar mass functions.  We first converted all stellar mass functions to a \citet{Chabrier2003} IMF, upon which all GASS derived quantities are based.  The stellar mass function from \citet{Bell2003_SMF} produces a total HIMF that agrees well with the HISMF we derived based on the \citet{Borch2006} stellar mass function.  The stellar mass function from \citet{Bernardi2010} yields an HISMF within 10\% of ours while \citet{Li2009} and \citet{Panter2007} yield HISMFs that are generally within 20\% of ours.  The HISMFs agree well in part because the stellar mass functions agree well in the GASS stellar mass range and are more discrepant at low (log M$_*$/M$_{\odot}$ $<$ 9.5) and high (log M$_*$/M$_{\odot}$ $>$ 11.0) stellar masses.

\section{Probabilities for Model Selection}

In Table \ref{tbl:prob} we list the probabilities discussed in Section \ref{sec:modelselection} for each of the six models and six stellar mass bins.  Column 3 lists the probability of the MCMC iteration with the highest probability.  Column 4 presents the Bayesian Information Criterion \citep[BIC;][]{Schwarz1978}.  The equation for the BIC is:

\begin{equation}
BIC = -2\times ln(P) + k ln(n)
\end{equation}

where P is the MCMC probability, k is the number of free parameters for each model (in our case, k=2 for the full and bent models and k=3 for the broken models), and n is the number of galaxies contributing to each fit.  A lower BIC indicates a better fit.

In Table \ref{tbl:prob} we also show the same sets of probabilities for the broken Schechter fits to our simulated data.  These probabilities are comparable to those derived from the observed data and show that the magnitude of the probabilities is strongly dependent on the number of galaxies in each stellar mass bin (the 11.25 $<$ log M$_*$/M$_\odot$) $<$ 11.5 bin contains many fewer galaxies). 

In Table \ref{tbl:probtot} we compare the total HISMF derived from each of the eight models to the data at log M$_{HI}$/M$_{\odot}$ $>$ 10.  We calculate the $\chi^2$ statistic as a discrepancy measure.  We also use the Poisson probability function to calculate the probability of the number of observed galaxies in each HI mass-stellar mass bin given the number of galaxies that each model predicts.

\begin{deluxetable}{crccccc}  
\tabletypesize{\scriptsize}
\tablecolumns{7}
\tablecaption{Model Probabilities\tablenotemark{a}}
\tablewidth{0pt}
\tablehead{
\colhead{Model} & \colhead{log M$_{*}$ = 10.0, 10.25} & \colhead{10.25, 10.5} & \colhead{10.5, 10.75} & \colhead{10.75, 11.0} & \colhead{11.0, 11.25} & \colhead{11.25, 11.5} }
\startdata
Schechter Full    &-81.61	&-82.61	&-86.59	&-78.10	&-79.42	&-17.92\\
Schechter Broken	&-80.75	&-82.13	&-83.08	&-74.31	&-77.87	&-18.11\\
Schechter Bent	&-80.78	&-82.21	&-83.37	&-74.86	&-78.45	&-18.14\\
Log-normal Full	&-87.27	&-86.69	&-94.60	&-84.12	&-82.65	&-18.28\\
Log-normal Broken	&-82.03	&-82.71	&-83.93	&-74.64	&-77.06	&-18.81\\
Log-normal Bent	&-84.29	&-83.24	&-86.80	&-76.95	&-77.07	&-18.55\\
Schechter Broken (simulation) &-76.97 &-76.56 &-75.97&-64.98&-66.83&-14.08\\
\tableline
Schechter Full	&172.11	&174.07	&182.31	&165.29	&167.97	&42.27\\
Schechter Broken	&174.82	&177.55	&179.86	&162.25	&169.43	&45.88\\
Schechter Bent	&170.45	&173.29	&175.87	&158.80	&166.02	&42.72\\
Log-normal Full	&183.43	&182.24	&198.33	&177.32	&174.43	&43.01\\
Log-normal Broken	&177.38	&178.71	&181.56	&162.91	&167.81	&47.28\\
Log-normal Bent	&177.47	&175.34	&182.74	&162.98	&163.26	&43.54\\
Schechter Broken (simulation) &167.27 &166.41 &165.63 &143.59 &147.35 &37.82
\enddata
\tablenotetext{a}{Log probabilities from MCMC runs (top seven lines) and BIC (bottom seven lines).}
\label{tbl:prob}
\end{deluxetable}

\begin{deluxetable}{ccc}  
\tabletypesize{\scriptsize}
\tablecolumns{3}
\tablecaption{Probabilities for Total HISMF at log M$_{HI}$ $>$ 10}
\tablewidth{0pt}
\tablehead{
\colhead{(1)} & \colhead{(2)} \\
\colhead{log M$_*$} & \colhead{$\chi^2$} & \colhead{ln(Poisson Probability)} }
\startdata
Schechter Full    &0.92	&-4.08\\
Schechter Broken	&0.99	&-3.89\\
Schechter Bent	&1.24	&-3.73\\
Log-normal Full    &5.14	&-16.50\\
Log-normal Broken	&3.14	&-10.42\\
Log-normal Bent	&2.59	&-9.83\\
Bivariate    &1.83	&-4.22\\
Trivariate	&2.38	&-4.03\\
\enddata
\label{tbl:probtot}
\end{deluxetable}


\end{document}